\documentclass[preprint,12pt,numbers,square]{elsarticle}
\usepackage{tabularray}
\UseTblrLibrary{booktabs}

\usepackage{amssymb}
\usepackage{amsmath}
\usepackage{comment}
\usepackage{todonotes}
\usepackage{blindtext}
\usepackage{epsfig}
\usepackage{multirow}
\usepackage{relsize}
\usepackage{multicol}
\usepackage{colortbl}
\usepackage{booktabs}

\newcommand{\Eq}[1]{Eq.~(\ref{eq:#1})}
\newcommand{\eq}[1]{\Eq{#1}}
\newcommand{\fig}[1]{Fig.~\ref{fig:#1}}
\newcommand{\tab}[1]{Tab.~\ref{tab:#1}}

\newcommand{\sect}[1]{Section~\ref{sec:#1}}

\usepackage{hyperref}
\usepackage{xspace}
\newcommand{\latinphrase}[1]{\textit{#1}}  
\newcommand{\etal}{\latinphrase{et~al.}\xspace}
\newcommand{\ie}{\latinphrase{i.e.}\xspace}

\usepackage[a4paper, margin=2.5cm]{geometry}
\usepackage[normalem]{ulem}
\usepackage{xcolor}
\newcommand{\camera}[1]{#1}

\newcommand{\update}[1]{{\color{black} #1}}  
\newcommand{\updateSecond}[1]{{\color{black} #1}} 

\journal{Knowledge-Based Systems}

\begin{document}

\begin{frontmatter}



\title{Continuous and complete liver vessel segmentation\\ with graph-attention guided diffusion}


\author{Xiaotong Zhang, Alexander Broersen, Gonnie CM van Erp, \\Silvia L. Pintea, Jouke Dijkstra*} 

\affiliation{organization={Radiology department, Leiden University Medical Center},
            addressline={Albinusdreef 2}, 
            city={Leiden},
            postcode={2333 ZA}, 
            state={South Holland},
            country={The Netherlands}}


\begin{abstract}
Improving connectivity and completeness are the most challenging aspects of liver vessel segmentation, especially for small vessels. 
These challenges require both learning the continuous vessel geometry, and focusing on small vessel detection.
However, current methods do not explicitly address these two aspects and cannot generalize well when constrained by inconsistent annotations.
Here, we take advantage of the generalization of the diffusion model and explicitly integrate connectivity and completeness in our diffusion-based segmentation model.
Specifically, we use a graph-attention module that adds knowledge about vessel geometry, and thus adds continuity.
Additionally, we perform the graph-attention at multiple-scales, thus focusing on small liver vessels. 
Our method outperforms \updateSecond{eight} state-of-the-art medical segmentation methods on two public datasets: \textsl{3D-ircadb-01} and \textsl{LiVS}. 
\camera{Our code is available at {\small \href{https://github.com/ZhangXiaotong015/GATSegDiff}{https://github.com/ZhangXiaotong015/GATSegDiff}}.}
\end{abstract}




\begin{keyword}


Liver vessel segmentation\sep CT \sep Diffusion \sep Multi-scale graph-attention
\end{keyword}

\end{frontmatter}



\section{Introduction}
\label{sec:introduction}

Liver cancer is the fourth leading cause of death according to statistics on cancer-related mortality \cite{b1}. 
Furthermore, the liver is a frequent site for metastasis of various primary tumors, such as gastrointestinal tumors, breast cancer, lung cancer, and melanoma \cite{b60}.  
Both primary and secondary liver cancer have multiple treatment options, including surgery, and various interventional oncology liver treatments. 
The preoperative planning of these treatments can be improved with accurate segmentation of the liver vessels \cite{b2,b58}.
In preoperative planning of liver tumor resection \cite{b58}, visualizing the spatial location between liver vessels and tumors in a $3$D view is essential to reduce surgical risk. 
Liver vessel segmentation is used primarily to ensure that the main vessels are not located near the planned resection region, thus reducing bleeding.
Liver vessels also indicate the boundaries for the Couinaud classification \cite{b58}.
Furthermore, it can assist in targeting the correct tumor nutrient supply vessel to decrease the recurrence rate in embolic therapies \cite{b4}. 
Hence, accurate liver vessel segmentation is essential in liver tumor surgery.
However, acquiring automatic liver vessel segmentation is challenging due to the complex anatomy.

\begin{figure}[t]
  \centering
  \includegraphics[width=.85\linewidth]{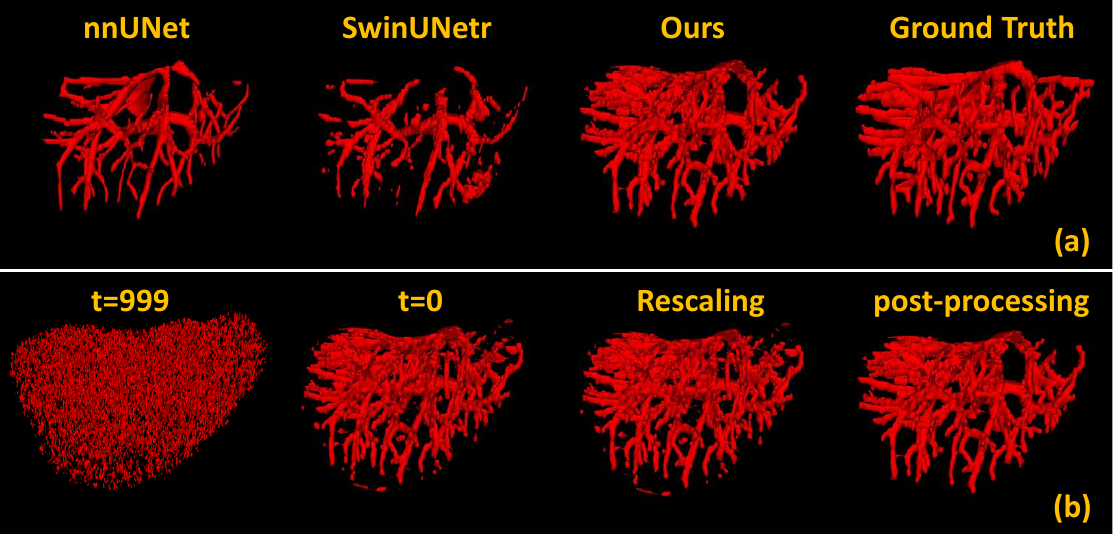}
  \caption{
    (a) Vessel trees of \textsl{nnUNet} \cite{b27}, \textsl{Swin UNETR} \cite{b28}, our proposed model and the ground truth. 
    (b) Vessel trees predicted by our method across different steps ($t{=}0$ is the final diffusion iteration). 
    Rescaling is the process of recovering the resolution of resized CT images to that of the original CT images.
    Our method predicts complete and continuous vessel trees. 
  }
  \label{fig:intro}
\end{figure}

Automatic liver vessel segmentation is performed on computed tomography (CT) images.
Traditionally, methods relied on image filtering \cite{b5,b6}, active contour models \cite{b7,b8}, or tracking methods \cite{b9,b11}. 
Currently, the leading methods on liver vessel segmentation are based on deep artificial networks \cite{b13,b14,b27,b28}. 
\textsl{nnUNet} \cite{b27} can extract features automatically from the CT images. 
However, \textsl{nnUNet} cannot ensure vessel continuity, as shown in \fig{intro}(a).
Attention-based methods \cite{b16,b18} as used in \textsl{Swin UNETR} \cite{b28} improve vessel continuity.
Yet, \textsl{Swin UNETR} cannot ensure completeness, by struggles with horizontally distributed vessel, as seen in \fig{intro}(a). 
In addition, the performance of the above Convolutional Neural Network (CNN) cannot generalize well, especially when the label annotation styles across data are different \cite{b61}. 
Thus, ensuring vessel continuity in all directions, localizing small vessels, and improving model generalization for the inconsistent annotation style, remains challenging.

\update{
Here, we address these challenges by proposing a model that makes continuous, and complete predictions, and generalizes well to different annotation styles.
The inherent anatomical structure of the vascular tree inspires our network design. 
We are motivated by the assumption that introducing an explicit vascular graph can help the model better capture irregular and long-range vessel connectivity, thus adding continuity.
Additionally, we use multiple scales in our graph structure, thus enabling the detection of small vessels, adding completeness to the model.
Moreover, prior work tends to overlook the underlying data distribution, resulting in a strong dependence on annotation quality.
To achieve accurate segmentation despite imperfect annotations, we rely on a diffusion model to learn the underlying data distribution, and mitigate the dependence on annotation quality.
}

Concretely, our model starts from a 2$D$ diffusion model \cite{b31,b32}. 
We opt for a $2$D rather than a $3$D diffusion model to reduce computational requirements.
To ensure vessel continuity, we add graph-attention layers \cite{b30} into the diffusion model. 
Because the graph is sparse, we compensate for this by integrating neighboring features on the graph in a local ensemble module \cite{b33}. 
The local ensemble module ensures a smooth transition between different nodes \cite{b33}. 
Secondly, to segment small vessels, we extract features at multiple scales in the nodes of the graph.
The effectiveness of these components is shown in \fig{intro}(b).

Our contributions are: 
(i) explicitly incorporating vessel continuity by adding graph-attention conditioning to a diffusion model for liver vessel segmentation;
(ii) explicitly focusing on small vessels by relying on multi-scale graph-features when conditioning the diffusion model;
(iii) continuous and complete vessel segmentations on two public datasets \textsl{3D-ircadb-01} \cite{b40} and \textsl{LiVS} \cite{b41}, compared to existing work.

\section{Related Work}
\noindent\textbf{Liver vessel segmentation.}
Liver vessel segmentation currently relies on \textsl{CNN} and attention methods. 
CNN methods either rely on \textsl{FCN} (fully convolutional networks) \cite{b14,b43}, or follow the \textsl{UNet} architecture \cite{b13,b15}. 
Attention methods can be grouped into self-attention \cite{b16,b48} and graph-attention \cite{b17,b18} methods.
\textsl{UNet} \cite{b12} and its variants are the effective for medical segmentation.
However, they struggle with imbalanced data \cite{b46}, such as between the vascular and the liver region.
Moreover, their performance is limited by the receptive-field size of the convolutions \cite{b47}. 
While the data imbalance can be mitigated by improving the \textsl{Dice} loss\cite{b13}, learning long-range dependencies remains a challenge \cite{b28}. 
To address this, \textsl{Swin-transformer} \cite{b34} uses a transformer architecture, and thus its receptive-field size covers the full input resolution.
Similarly, graph-attention networks \cite{b29} capture long-range dependencies by computing node attention-coefficients.
Augmenting the \textsl{UNet} by \textsl{Swin-transformer} \cite{b16}, or using the graph-attention to assist the \textsl{UNet} training\cite{b18} is also effective in practice. 
\update{In addition to methods specifically developed for liver vessel segmentation, several works on retinal vessel segmentation \cite{b64}, skin lesion segmentation \cite{b66}, and general medical image segmentation \cite{b65,b67} also incorporate self-attention mechanisms or U-Net variants to improve segmentation accuracy.}
Here, we also combine the capabilities of \textsl{CNNs} and attention mechanisms.
Moreover, we jointly enforce continuity of the vessel-tree segmentations and focus on small vessels.

\medskip\noindent\textbf{Diffusion models for medical image segmentation.}
Diffusion \cite{b31} methods showcase promising results for medical image segmentation. 
These diffusion methods are either non-dynamic conditioning \cite{b19} or dynamic conditioning \cite{b20,b25}.
The non-dynamic models concatenate medical images to the input, and do not adapt this conditioning information over time. 
Their performance \cite{b19,b24} is comparable to (or lower than) \textsl{nnUNet} \cite{b27}, which is the standard medical segmentation baseline. 
On the other hand, dynamic conditioning methods \cite{b25} use an extra encoder to generate time-dependent conditioning information.
Similarly to the non-dynamic conditioning models, their accuracy is limited.
More recently, \update{\textsl{HiDiff} \cite{b63}} and \textsl{MedSegDiff} \cite{b20} using a hybrid constrained method, obtains state-of-the-art results. 
\textsl{MedSegDiff} \cite{b20} predicts a weighted combination of a diffusion segmentation and an auxiliary segmentation from the conditioning branch. 
\update{\textsl{HiDiff} \cite{b63} also relies on a prior segmentation as one of the conditioning inputs to guide the diffusion model.}
This combination weakens the contribution of the diffusion model.
Here, we build on a $2$D dynamic conditioning diffusion model constrained by a hybrid loss, yet we only predict the diffusion segmentation.

\medskip\noindent\textbf{Graph-based methods for medical image segmentation.}
Graphs are one of the most intuitive way to represent complex anatomical structures. 
Graph-based methods can segment medical tree structures \cite{b50,b36}, but also non-structural medical data \cite{b52,b54}. 
Although, the graphs add long-range dependencies in CNNs, they tend to miss small branches \cite{b55}. 
This is due to the sparsity of the nodes, causing loss of information. 
Here, we also rely on graphs to add connectivity for vessel segmentation. 
To avoid missing small vessels, we use the local ensemble module of LIIF \cite{b33} which smoothes the feature between nodes.
Additionally, we use multiscale graph features to focus on small vessels.


\begin{figure*}
    \centering
    \includegraphics[width=.9\textwidth]{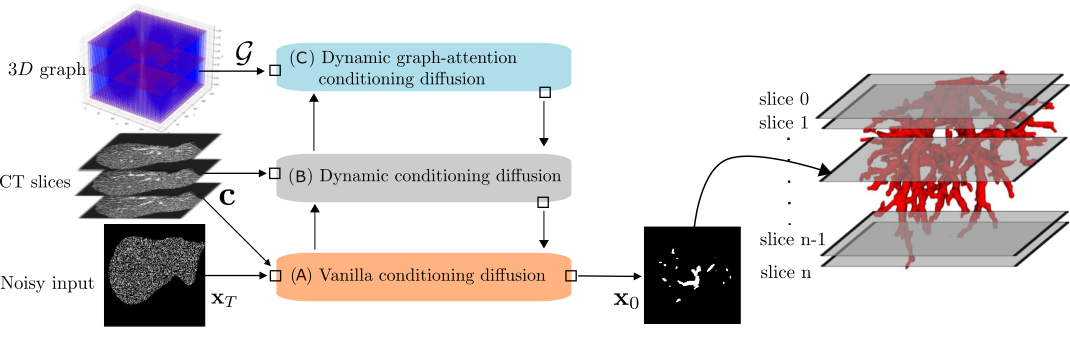}
    \caption{
    \textbf{Overview of our model:}
    (A) A vanilla diffusion model starting from noisy images $\mathbf{x}_T$, and predicting segmentation masks $\mathbf{x}_0$ (in orange);
    (B) A dynamic conditioning model, conditioned on three CT slices $\mathbf{c}$ (in gray); and
    (C) A multiscale graph-attention conditioning model, starting from a graph structure $\mathcal{G}$ (in blue).
    }
    \label{fig:overview}
\end{figure*}

\section{Diffusion conditioning models}
Our model contains three components, as show in \fig{overview}:
(A) The vanilla conditioning diffusion model over three CT slices, for conditioning; 
(B) Dynamic conditioning diffusion, starting from the same CT slices but using a separate encoder; and
(C) Conditioning diffusion model with multiscale graph-attention guidance.

\subsection{(A) Vanilla conditioning diffusion model}
\noindent\textbf{Diffusion model.}
Conditioning diffusion models extend the Denoising Diffusion Probabilistic Models (DDPM) \cite{b31}.
DDPM is composed of a forward process and a reverse process.

The forward process gradually adds Gaussian noise to the inputs $\mathbf{x}_0$ over a number of $T$ timesteps.
In our case $\mathbf{x}_0$ is the ground truth vessel segmentation mask.
The variance of the Gaussian noise is modeled by $\beta_{t}$, which is typically a linear function of $t$.
Thus, the distribution of the noisy vessel mask $\mathbf{x}_{t}$ given the ground truth $\mathbf{x}_{0}$, is:
\begin{align}
    q(\mathbf{x}_t | \mathbf{x}_0) = \mathcal{N}(\mathbf{x}_t; \sqrt{\bar{\alpha}_t} \mathbf{x}_0, (1 - \bar{\alpha}_t)I),
    \label{eq:1}
\end{align} 
where $\bar{\alpha}_t = \prod_{s=1}^{t} (1-\beta_s)$. 
In the forward process, we obtain the noisy vessel mask $\mathbf{x}_{t}$ from $\mathbf{x}_{0}$ as a linear combination with the noise $\epsilon_t \sim \mathcal{N}(0, I)$:
\begin{align}
    \mathbf{x}_t = \sqrt{\bar{\alpha}_t} \mathbf{x}_0 + \sqrt{1-\bar{\alpha}_t} \epsilon_t.
    \label{eq:2}
\end{align} 

In the reverse process, we train a model $p_\theta$ with parameters $\theta$, to iteratively denoise an input noisy image $\mathbf{x}_T$. 
This aims to recover the clean segmentation mask $\mathbf{x}_0$. The model $p_\theta$ follows a Gaussian distribution:
\begin{align}
    p_\theta(\mathbf{x}_{t-1} | \mathbf{x}_t) = \mathcal{N}(\mathbf{x}_{t-1}; \mu_\theta(\mathbf{x}_t,t), \Sigma_\theta (\mathbf{x}_t, t)).
    \label{eq:3}
\end{align} 
The mean $\mu_{\theta}(\mathbf{x}_t, t)$ and variance $\Sigma_{\theta}(\mathbf{x}_t, t)$ of the reverse process are functions of a noise model $\epsilon_\theta (\mathbf{x}_t, t)$. 
The noise $\epsilon_\theta$ is typically modelled by a \textsl{UNet} \cite{b12,b13}. 
This \textsl{UNet} noise model $\epsilon_{\theta}$ is trained by minimizing the difference between the estimated noise $\epsilon_{\theta}$ and the true noise $\epsilon_t$ at a number of sampled timesteps $t{\sim}[1,T]$:
\begin{align}    
    L_{\text{den}}(\mathbf{x}_0, \theta) = \mathbb{E}_{t\sim[1,T],\mathbf{x}_0,\epsilon_t} \lVert \epsilon_t - \epsilon_{\theta}(\mathbf{x}_t, t) \rVert^2.
    \label{eq:4}
\end{align}

\medskip\noindent\textbf{Conditioning diffusion models.}
Following \cite{b19}, we add a conditioning to the DDPM model that is independent of the timestep $t$. 
We use three consecutive CT slices, $\mathbf{c}$, as conditioning for our DDPM model. 
We evaluate the vanilla conditioning model in the experiments.

\begin{figure*}[t]
    \centering
    \includegraphics[width=1\textwidth]{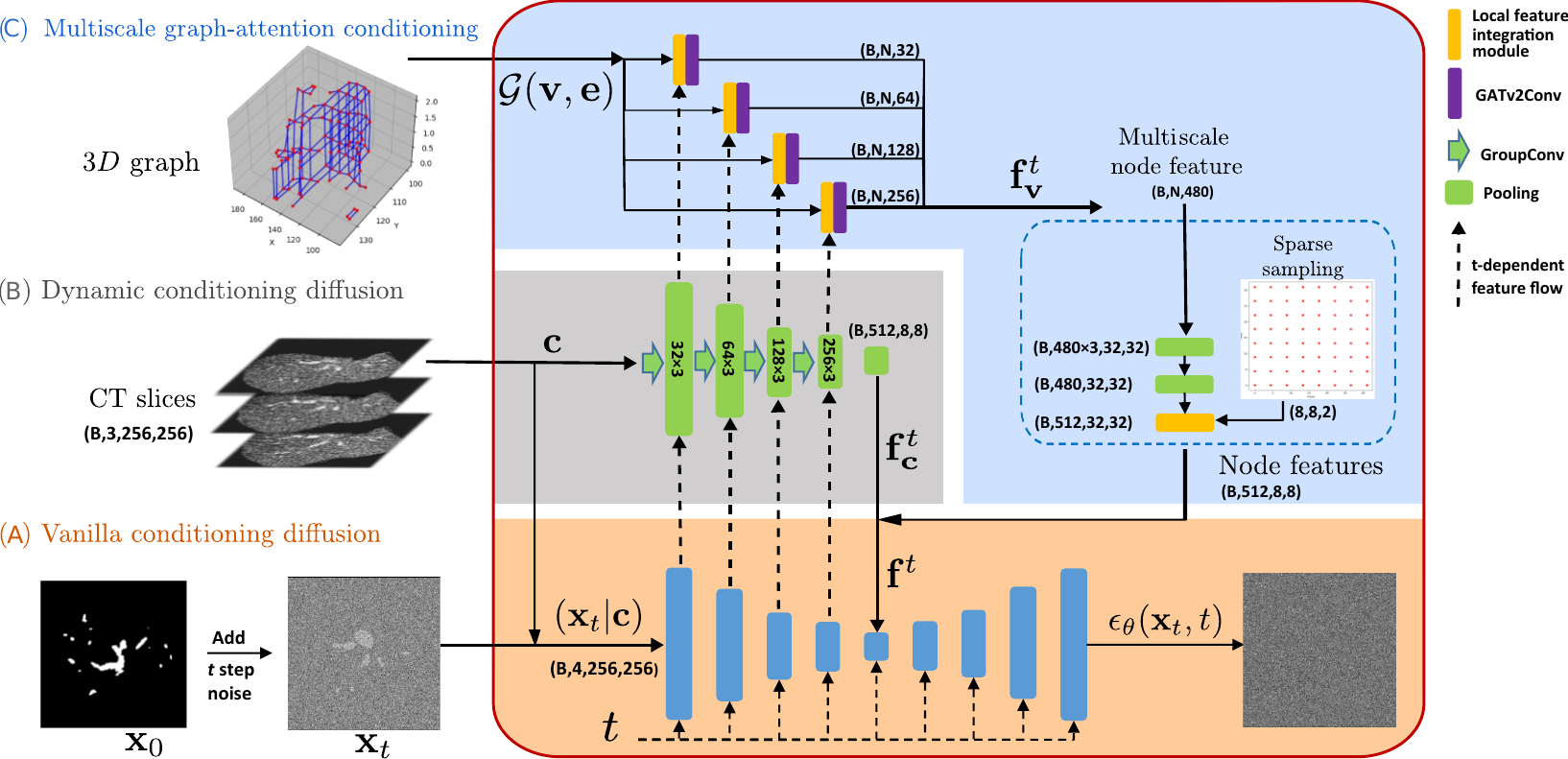}
    \caption{
    \textbf{Our network architecture:}
    (A) vanilla conditioning diffusion (orange);
    (B) dynamic conditioning diffusion (gray); 
    (C) multiscale graph-attention conditioning (blue).
    These components interact through the vertical dashed\slash solid arrows.
    The dashed upwards arrows adapt the conditioning over time.
    The solid downwards arrows add the conditioning features:
    CT slices $\mathbf{c}$, CT slice embeddings $\mathbf{f}_\mathbf{c}^t$, and graph embeddings $\mathbf{f}_\mathbf{v}^t$. 
    }
    \label{fig:detailed}
\end{figure*}

\subsection{(B) Dynamic conditioning model}
Vanilla conditioning cannot adapt the condition across the noise levels (time steps).
To address this, at every diffusion timestep $t$, we embed the CT slices, $\mathbf{c}$, by using the \textsl{GenericUNet} encoder from \textsl{nnUNet} \cite{b27}, giving rise to  $\mathbf{f}_\mathbf{c}^t$.
To obtain time-dependent conditioning, $\mathbf{f}_\mathbf{c}^t$, at each timestep $t$ into the bottleneck of the \textsl{vanilla conditioning diffusion model}, as shown in \fig{detailed}(ii) (solid downward arrow).
Additionally, we use group convolutions in the conditioning, thus keeping the features per slice separate.
To let the CT embedding $\mathbf{f}_\mathbf{c}^t$ adapt over time, we merge the noisy features of the \textsl{vanilla conditioning model} into the \textsl{dynamic conditioning model}, at the appropriate depth (in \fig{detailed}(ii) with dashed arrows). 

\subsection{(C) Multiscale graph-attention conditioning model}
The \textsl{vanilla conditioning} and the \textsl{dynamic conditioning} both use the denoising loss in \eq{4}. 
To make use of the geometric structure of vessels, we first map the $3$D vessel tree into a graph $\mathcal{G}$.
Subsequently, we use graph-attention \cite{b30} to add this geometric structure as a condition into the diffusion model. 

\medskip\noindent\textbf{Vessel graph construction.}
We construct a $3$D vessel graph $\mathcal{G}{=}(\mathbf{V}, \mathbf{E})$, where the nodes $\mathbf{V}$ are locations along the vessel, and the edges $\mathbf{E}$ indicate vascular connectivity. 
We start from the full volume $[D{\times}H{\times}W]$ of a ground truth vessel tree, and we split it into non-overlapping sub-volumes $[d{\times}h{\times}w]$. 
Each node $\mathbf{v}{\in}\mathbf{V}$ corresponds to a sub-volume and is the average of the voxel coordinates along the vessel region. 
If there is no vascular annotation in a sub-volume, we use the central voxel as the node.
The graph edges, $\mathbf{e}{\in}\mathbf{E}$, are the geodesic distances between nodes, as in the Vessel Graph Network of Shin et al. \cite{b36}. 
Nodes with a small distance, but belonging to different vessel branches, should not be connected. 
For this, we use the binary vessel label as a speed function to calculate the travel time from one node to another, as in Li et al. \cite{b18}.

\medskip\noindent\textbf{Graph training and inference.}
During training, we construct the graph $\mathcal{G}$ using the ground truth vessel masks, as in \fig{detailed}(C).
We only display the foreground nodes, corresponding to the location of the vessels, and leave the background nodes transparent.
During inference, we do not have access to the ground truth vessel masks.
Therefore, we input to the \textsl{multiscale graph-attention conditioning model} a fully-connected graph, as in \fig{overview}(C).
This is a viable choice, because during training, the graph helps adapt the weights of the component (C).
Specifically, via the graph-attention, it extracts informative features from component (B). 
At inference, these attention-weights are trained, and can be applied on the new input features coming from component (B). 
In the ablation studies (\sect{abl}) we show the effectiveness of using a fully-connected graph at inference.

\medskip\noindent\textbf{Multiscale graph-attention.}
We use the CT-slice embeddings, $\mathbf{f}_\mathbf{c}^t$, from the \textsl{dynamic conditioning model} to extract node features at each timestep $t$ (dashed arrows in \fig{detailed}(C)).
We process the node features via a graph-attention layer GATv2 \cite{b30} and a local feature integration module (LIIF) \cite{b33}, to obtain node attention-coefficients.
We concatenate these node attention-coefficients over the different scales (network depths), coming from the \textsl{dynamic conditioning model}, giving rise to multiscale node attention-coefficients: $\mathbf{f}_\mathbf{v}^t$ at each node $\mathbf{v}$ and timestep $t$.

The nodes in the vessel graph $\mathcal{G}$ are sparse (only $32{\times}32$ nodes for one CT slice).
To compensate for this sparsity, we use the local features integration module LIIF \cite{b33}, which is popular for image super-resolution.
We extend LIIF from $2$D to $3$D, and apply it on our CT-embeddings $\mathbf{f}_\mathbf{c}^t$.
Specifically, for a CT-slice embedding $\mathbf{f}_\mathbf{c}^{(t,i)}$ at timepste $t$ and location $i$ corresponding to a graph node $\mathbf{v}_i$, we use the graph neighboring locations $\mathbf{v}_{ne(i)}$ to define a new embedding:
\updateSecond{\begin{align}
    \hat{\mathbf{f}}_\mathbf{c}^{(t,i)} &= \sum_{ne(i)} \frac{\mathcal{S}(\mathbf{v}_i,\mathbf{v}_{ne(i)})}{\mathcal{\overline{S}} } ~\text{LFI}\left(\mathbf{f}_\mathbf{c}^{(t,ne(i))}, \mathbf{v}_i - \mathbf{v}_{ne(i)}\right),
    \label{eq:5}
\end{align}}
where our neighboring locations $ne(i)$ vary across $x,y$, and $z$ directions, rather than just $x,y$; 
and $\mathcal{S}(\cdot,\cdot)$ computes the area a 3D cube between the graph node $\mathbf{v}_i$ and its neighbor $\mathbf{v}_{ne(i)}$; 
\updateSecond{$\mathcal{\overline{S}}{=}\sum_{ne(i)}\mathcal{S}(\mathbf{v}_i,\mathbf{v}_{ne(i)})$}; 
\updateSecond{$\mathbf{f}_\mathbf{c}^{(t,ne(i))}{=}\text{GridSample}(\mathbf{f}_\mathbf{c}^{t},ne(i))$} \cite{b69};
and the local feature integration module is \updateSecond{$\text{LFI}\left(\mathbf{f}_\mathbf{c}^{(t,ne(i))}, \mathbf{v}_i {-} \mathbf{v}_{ne(i)}\right){=}\text{Conv2d}\left(
\text{Cat}\left(\mathbf{f}_\mathbf{c}^{(t,ne(i))}, \mathbf{v}_i {-} \mathbf{v}_{ne(i)}\right)\right)$}\cite{b33}. 
\updateSecond{Inside the function $\text{LFI}(\cdot,\cdot)$, we extract the neighboring features of each node based on their 3D coordinates $\mathbf{v}_{ne(i)}$, and embed the relative position differences $(\mathbf{v}_i {-} \mathbf{v}_{ne(i)})$, which represent the spatial relationships among neighborhood features.}

Subsequently, we use new CT-slice embeddings $\hat{\mathbf{f}}_\mathbf{c}^{(t,i)}$ to extract node features via a graph-attention layer GATv2 \cite{b30}, for every two neighboring locations $i$, $j$:
{\small
\begin{align}
    \mathbf{f}_\mathbf{v}^{(t,i,j)} = \frac{\exp\left[\mathbf{a}^\top \text{LeakyReLU}\left(\mathbf{W} (\hat{\mathbf{f}}_\mathbf{c}^{(t,i)} + \hat{\mathbf{f}}_\mathbf{c}^{(t,j)})\right)\right]}{\sum_{j^\prime} 
    \exp\left[\mathbf{a}^\top \text{LeakyReLU}\left(\mathbf{W} (\hat{\mathbf{f}}_\mathbf{c}^{(t,i)} + \hat{\mathbf{f}}_\mathbf{c}^{(t,j^\prime)})\right)\right]}.
    \label{eq:6}
\end{align}}

During training, we want to learn which CT embeddings correspond to the foreground vessels and which not.
Therefore, we process the node features via a convolutional layer with $\text{sigmoid}$ activation, to obtain $\hat{\mathbf{f}}^t_\mathbf{v}{=}\text{sigmoid}(\text{Conv}(\mathbf{f}^t_\mathbf{v}))$, and optimize the model parameters $\theta$ using a binary cross-entropy loss:
\begin{align}
    L_{\text{graph}}(\mathcal{G},\theta) &= -\sum_\mathbf{v \in \mathcal{G}}\sum_{t} \log(\hat{\mathbf{f}}^t_\mathbf{v}).
    \label{eq:7}
\end{align}

\subsection{Overall diffusion model conditioning} 
We use the multiscale graph-attention embeddings $\mathbf{f}_\mathbf{v}^t$ together with the CT embeddings, $\mathbf{f}_{\mathbf{c}}^t$, to condition the reverse diffusion process.
Therefore, \eq{3} defining the reverse process, becomes:
\begin{align}
    p_{\theta}(\mathbf{x}_{t-1} | \mathbf{x}_t) &= \mathcal{N}(\mathbf{x}_{t-1}; \mu_{\theta}(\mathbf{x}_t | \mathbf{f}^t, t), \Sigma_{\theta}(\mathbf{x}_t | \mathbf{f}^t, t)),\\
    \text{ where }\mathbf{f}^t &= \mathbf{f}_{\mathbf{c}}^t + \mathbf{f}_{\mathbf{v}}^t.
    \label{eq:8}
\end{align}

\subsection{Overall loss function}
Our overall loss function, used to fit the model parameters $\theta$, is a combination of the denoising loss in \eq{4} and graph loss in \eq{7}:
\begin{align}
    L_\text{total}(\mathbf{x}_0, \mathbf{c}, \mathcal{G}, \theta) &= L_{\text{den}}(\mathbf{x}_0, \mathbf{c}, \mathcal{G}, \theta) + L_{\text{graph}}(\mathcal{G},\theta),
    \label{eq:9}
\end{align}
where the denoising loss $L_\text{den}$ relies on the reverse diffusion process in \eq{8}.


\section{Experimental analysis}
\noindent\textbf{Liver vessel segmentation datasets.}
We use two public datasets: \textsl{3D-ircadb-01} \cite{b40} and \textsl{LiVS} \cite{b41}, as detailed in \tab{datasets}.
\textsl{3D-ircadb-01} contains $20$ cases, while \textsl{LiVS} contains $532$ cases.
In the \textsl{3D-ircadb-01} dataset every slice is annotated, but some small vessels are not annotated \cite{b15}.
In the \textsl{LiVS} dataset only a subset of randomly chosen slices are annotated. 
\begin{table}[b]
    \caption{Dataset overview. 
    In \textsl{3D-ircadb-01} \cite{b40} every CT slice is annotated, but some small vessels are missing with an inconsistent annotation style.
    \textsl{LiVS} \cite{b41} contains more CT volumes and has stable annotation style, but only a subset of the slices are annotated.
    }
    \centering 
    \resizebox{.7\linewidth}{!}{%
    \begin{tabular}{lll}
        \toprule
        & \textsl{3D-ircadb-01} & \textsl{LiVS} \\ \midrule
        Available scans     &   20      &   532 \\
        Used scans          &   20      &   303 \\
        \multirow{2}{*}{Exclusion} & completeness score & \#annotations (${<3}0$ slices)\\
                            & CT thickness ${\geq}2.5$ mm & CT thickness ${\geq}2.5$mm\\
        Pixel spacing       & $0.57$ mm -- $0.87$mm & $0.51$ mm -- $0.98$ mm\\
        Slice thickness     & $1.00$ mm -- $4.00$mm & $0.62$ mm -- $5.00$ mm\\
        Continuous annotation &Yes                  &       No \\
        High-contrast tumors     &No                &       Yes\\
        Annotation consistency  &Low                &       High\\
        \bottomrule
    \end{tabular}}
    \label{tab:datasets}
\end{table}

CT scans with thick slices (${\geq}2.5$ mm for \textsl{3D-ircadb-01} and ${\geq}5$ mm for \textsl{LiVS}) are rare, forming outliers. 
Thus, for \textsl{LiVS} we exclude these cases from training and test. 
For the relatively small \textsl{3D-ircadb-01} dataset, we only exclude the cases with thick slices from the testset. 
To effectively evaluate on the small \textsl{3D-ircadb-01} dataset, we asked a clinical expert to score the completeness of the annotated vessels. 
After thickness exclusion, cases $\{04, 06, 08, 11, 16\}$ were marked as complete. 
Therefore, we use these for testing.
Specifically, we perform leave-one-out cross-validation on these five cases, where at every fold we train on 19 cases and test on 1 test sample, from the list above.  
We average the results over all folds.
For \textsl{LiVS} we report average metrics over 3-fold cross-validation.

\medskip\noindent\textbf{Data pre- and post-processing.}
For the \textsl{3D-ircadb-01} \cite{b40} dataset, we first crop the liver region and resize the cropped CT slices to $256{\times}256$ px. 
The liver masks exclude the \textsl{vena cava}. 
We clip the intensity of the CT slices to $[0, 400]$ HU (Hounsfield units). 
The CT slices are already cropped, resized, and clipped in the \textsl{LiVS} \cite{b41} dataset.
Because not all \textsl{LiVS} slices are annotated, we use ITK-SNAP\cite{b39} to interpolate the annotations. 
In our method, we sample the central slice of the $2.5$D block only from the set of CT slices with ground truth annotations. 

During inference, we rescale the diffusion predictions back to the physical resolution of the original CT image of $512{\times}512$ px.
During post-processing, we remove disconnected noisy spots, with a volume less than $1\%$ of the largest connected region, using connected region analysis \cite{b56}. 
When the vessel annotations are continuous in the longitudinal direction, we find post-processing \cite{b13} more effective to obtain the final vessels.
For discontinuous annotations, ensemble inference with different seeds, is more effective.

\medskip\noindent\textbf{Evaluation metrics.}
For all our experiments we report: Dice similarity coefficient (\textsl{DSC}), voxel-wise sensitivity (\textsl{Sen}), voxel-wise specificity (\textsl{Spe}) \cite{b42}, centerline Dice (\textsl{clDice}) \cite{b59}, and a custom connected region-wise connectivity (\textsl{Con}) following  Geg\'undez-Arias \etal \cite{b38}. 
The \textsl{Con} metric is the ratio of the total number of connected regions, in the predicted tree $\mathcal{T}$ and the total number of connected regions in the ground truth vessel tree $\mathcal{T}^*$:
\begin{align}
    \text{Con}(\mathcal{T},\mathcal{T}^*) = \frac{|\text{comp}(\mathcal{T})|}{|\text{comp}(\mathcal{T}^*)|}\geq1,
    \label{eq:10}
\end{align} 
where $\text{comp}(\cdot)$ computes the connected components. 
We only consider connected regions with a volume greater than $120$ mm$^3$, as in Huang \etal \cite{b13}. 
We exclude over-connected segmentations, where $\text{Con}{<}1.0$. 

\medskip\noindent\textbf{Baseline models.}
We compare our model with \update{seven} state-of-the-art medical segmentaiton methods, including \update{three} diffusion-based methods (\update{\textsl{HiDiff}\cite{b63}}, \textsl{MedSegDiff}\cite{b20}, \textsl{EnsemDiff}\cite{b19}), \update{two} self-attention methods \update{\textsl{MERIT}\cite{b62}} and \textsl{Swin UNETR} \cite{b28}, one self-configuring method \textsl{nnUNet} \cite{b27} and one specific liver vessel segmentation method \cite{b41}.
The \update{\textsl{HiDiff}}, \update{\textsl{MERIT}}, \textsl{MedSegDiff} and \textsl{EnsemDiff} use inputs of size $[3{\times}256{\times}256]$, while \textsl{Swin UNETR} and \textsl{nnUNet} are $3$D methods starting from the cropped CT slices as input.
\textsl{Swin UNETR}, \textsl{EnsemDiff} and \textsl{MedSegDiff} require ensembles. 
For \textsl{Swin UNETR}, we train 5 models and ensemble their segmented results. 
For \textsl{EnsemDiff} and \textsl{MedSegDiff}, we ensemble 5 diffusion results using different random seeds, but with a single training. 
Our method does not use ensembles when training with continuous annotations on \textsl{3D-ircadb-01}, but uses $5{\times}$ ensembles for discontinuous annotations on \textsl{LiVS}.

\medskip\noindent\textbf{Implementation details.}
We perform all our experiments on 1 NVIDIA RTXA6000 GPU with 48 GB memory. 
We use the \textsl{AdamW} optimizer with an initial learning rate of $1{\times}10^{-4}$ and a batch size of $10$. 
We input 4 channels: namely, three CT-slices and one noisy ground truth at each time step $t$ during training, and we use a random Gaussian noise channel during inference. 
Our model converges within $160$k iterations. 
We train the \textsl{EnsemDiff} and \textsl{MedSegDiff} models following their official implementation for $60$k\cite{b19} and $100$k\cite{b20}, respectively.
\update{We also train the \textsl{MERIT}\cite{b62} and \textsl{HiDiff}\cite{b63} using their official implementations and recommended configurations.}
On both the \textsl{3D-ircadb-01} and \textsl{LiVS} datasets, we use the standard DDPM sampling scheme \cite{b32} with $1000$ denoising steps, during inference, for all the diffusion-based experiments.
For our model, the CT block size is $[3{\times}256{\times}256]$ and the vessel graph $\mathcal{G}$ consists of nodes of size $(N,3)$ and edges of size $(E,2)$. 
Where we set the number of graph nodes $N$ is $32{\times}32$ per CT slice.

\subsection{Comparative experiments on \textsl{3D-ircadb-01} dataset}
\noindent\textbf{Quantitative evaluation on \textsl{3D-ircadb-01}.}
\tab{results1} provides the numerical evaluation of our proposed model, \update{\textsl{HiDiff} \cite{b63}, \textsl{MERIT} \cite{b62}}, \updateSecond{\textsl{TransUNet}} \cite{b68}, \textsl{MedSegDiff} \cite{b20}, \textsl{EnsemDiff} \cite{b19}, \textsl{Swin UNETR} \cite{b28}, \textsl{nnUNet} \cite{b27} and Gao et al. \cite{b41}.
We group the methods per network representation: $3$D (using $3$D convolutions) or $2.5$D (using $2$D convolutions over $2.5$D CT slices). 
The $3$D non-diffusion methods such as \textsl{nnUNet} and \textsl{Swin UNETR} are \update{comparable to or being} outperformed by \textsl{MedSegDiff} \update{and \textsl{HiDiff}}, but are better than \textsl{EnsemDiff}.
Intuitively, diffusion-based methods with dynamic conditioning like \textsl{MedSegDiff} \update{and \textsl{HiDiff}} perform better than the non-dynamic conditioning methods, like \textsl{EnsemDiff}. 
Although, \update{\textsl{HiDiff}}, \textsl{EnsemDiff}, \textsl{MedSegDiff} and our proposed model are built on diffusion models, our method still exceeds them in terms of \textsl{DSC} and \textsl{Sen} scores. 
Our model exceeds the \update{best-performing} baselines \update{for individual metrics} by \updateSecond{$1.49\%$} and \update{$6.61\%$} in \textsl{DSC} and \textsl{Sen} scores, as shown in \tab{results1}.
These improvements are due to the graph-attention conditioning, adding continuity and completeness to the vessel predictions.
Moreover, our method has the lowest standard deviation for \textsl{DSC} and \textsl{Sen} compared to the other methods.
This indicates that our model tends to make more stable predictions.

The \textsl{Spe} metric evaluates the degree of false positive predictions for a segmented liver vessel tree.
\textsl{Swin UNETR}, \updateSecond{\textsl{TransUNet}}, \textsl{EnsemDiff}, \update{\textsl{HiDiff}} and \textsl{MedSegDiff} obtain comparable \textsl{Spe} scores. 
However, our method obtains a slightly lower score than the others.
The lower \textsl{Spe} scores could be explained by the the missing small-vessel annotations. 
Interestingly, in \tab{results1}, \update{\textsl{MERIT} and} \textsl{nnUNet} have the highest averaged \textsl{Spe} but with lower averaged \textsl{Sen}, which is a trade-off between segmentation accuracy and completeness.

\tab{results1} also reports vessel connectivity, in the \textsl{clDice} \cite{b59} and \textsl{Con} metrics. 
\textsl{Con} (in \eq{10}) should ideally be as close as possible to $1$.
Additionally, we also report in the brackets the number of connected regions of the segmented vessel tree and the ground truth. 
Our method achieves the highest \textsl{clDice} score compared to the baselines, \updateSecond{with the exception of \textsl{TransUNet}}. 
\updateSecond{The \textsl{clDice} scores of both our method and \textsl{TransUNet} are ${\approx}75\%$}, which indicates that both methods can fit the centerlines of the segmented liver vessel tree, following the ground truth.
\textsl{nnUNet} achieves the \updateSecond{third} highest \textsl{clDice} score.
However, the \textsl{clDice} calculation depends on the erosion centerline, which can introduce bias.
The \textsl{Con} score of our method is the closest value to $1$ compared to the other methods, which is indicative of the continuous predictions in our model. 
Interestingly, \textsl{MedSegDiff} which is also based on a diffusion model, obtains the worst connectivity score.
This may be due to less precise auxiliary segmentations used in \textsl{MedSegDiff}.
\begin{table*}[t]
    \centering
    \caption{
    \textbf{Results on the \textsl{3D-ircadb-01} \cite{b40} dataset}.
    We compare our model with: \update{\textsl{HiDiff}}\cite{b63}, \update{\textsl{MERIT}}\cite{b62}, \updateSecond{\textsl{TransUNet}}\cite{b68}, \textsl{MedSegDiff}\cite{b20}, \textsl{EnsemDiff}\cite{b19}, \textsl{Swin UNETR}\cite{b28}, \textsl{nnUNet}\cite{b27} and \cite{b41}.
    Our method is the best in terms of \textsl{DSC}, \textsl{clDice}, \textsl{Sen} and \textsl{Con} scores, but worse in \textsl{Spe} scores. 
    Interestingly, \textsl{nnUNet} has the highest \textsl{Spe} score and the lowest \textsl{Sen} score, which may be due to a trade-off between detailed and accurate segmentation. 
    }
    \resizebox{1\linewidth}{!}{%
    \begin{booktabs}{lcllllll}
    \toprule
        & Type & \textsl{DSC} (\%) & \textsl{clDice} (\%) & \textsl{Sen} (\%) & \textsl{Spe} (\%) & \textsl{Con} ($\rightarrow 1$)\\
    \midrule
    Gao et al.\cite{b41}& 2D &$60.19\pm4.69$ &$56.48\pm11.74$ &$64.98\pm4.48$ &$99.80\pm0.10$ &$8.73\,(96/11)$\\ 
    \midrule
    \textsl{nnUNet}\cite{b27}& 3D&$60.54\pm6.86$ &$71.60\pm4.37$ &$44.76\pm8.09$ &$\textbf{99.99}\pm0.00$ &$2.36\,(26/11)$ \\
    \textsl{Swin UNETR}\cite{b28}& 3D&$55.61\pm9.80$  &$62.75\pm7.91$ &$42.30\pm11.46$ &$99.98\pm0.01$ & $5.45\,(60/11)$ \\
    \updateSecond{\textsl{TransUNet}}\cite{b68}& \updateSecond{3D}& \updateSecond{$69.77\pm3.73$} &\updateSecond{$\textbf{75.83}\pm5.50$} &\updateSecond{$58.63\pm11.70$} &\updateSecond{$99.97\pm0.01$} &\updateSecond{$3.73\,(41/11)$} \\
\midrule
    \textsl{EnsemDiff}\cite{b19}& 2.5D &$55.07\pm9.70$ &$60.99\pm9.64$ &$40.45\pm10.60$ &$99.98\pm0.02$ &$7.18\,(79/11)$ \\
    \textsl{MedSegDiff}\cite{b20}& 2.5D &$59.67\pm7.74$ &$66.02\pm7.87$  &$47.42\pm10.43$ &$99.95\pm0.05$ & $8.64\,(95/11)$ \\
    \update{\textsl{MERIT}}\cite{b62}& \update{2.5D} &\update{$49.50\pm9.57$} &\update{$53.00\pm8.79$} &\update{$34.30\pm8.81$} &\update{$99.99\pm0.01$} &\update{$5.50\,(59/11)$} \\
    \update{\textsl{HiDiff}}\cite{b63}& \update{2.5D} &\update{$63.32\pm4.01$} &\update{$60.24\pm2.71$} &\update{$54.45\pm7.25$} &\update{$99.94\pm0.01$} &\update{$6.70\,(72/11)$} \\
    Ours & 2.5D& $\textbf{71.26}\pm1.93$& $74.61\pm1.21$ & $\textbf{71.59}\pm4.07$& $99.89\pm0.04$& $\textbf{1.09}\,(12/11)$ \\
    \bottomrule
    \end{booktabs}}
    \label{tab:results1}
\end{table*}
\begin{figure*}[t]
  \centering
  \begin{tabular}{c@{\hskip 0.01in}c@{\hskip 0.01in}c@{\hskip 0.01in}c@{\hskip 0.01in}c@{\hskip 0.01in}c@{\hskip 0.01in}c@{\hskip 0.01in}c@{\hskip 0.01in}c@{\hskip 0.01in}c}
    \includegraphics[width=0.1\textwidth]{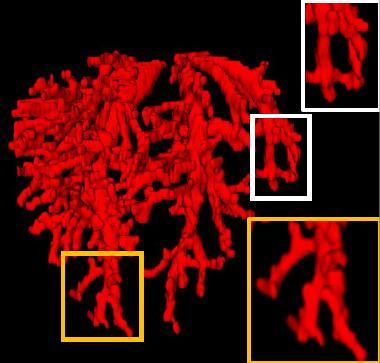} &
    \includegraphics[width=0.1\textwidth]{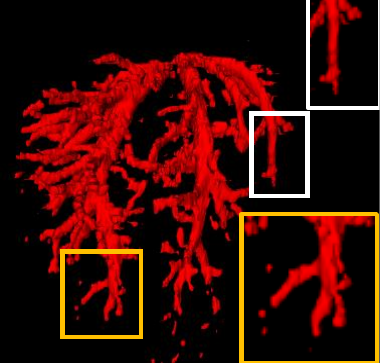}&
    \includegraphics[width=0.1\textwidth]{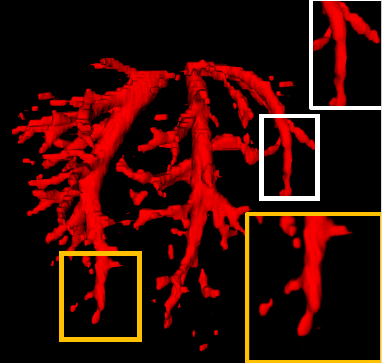}&
    \includegraphics[width=0.1\textwidth]
    {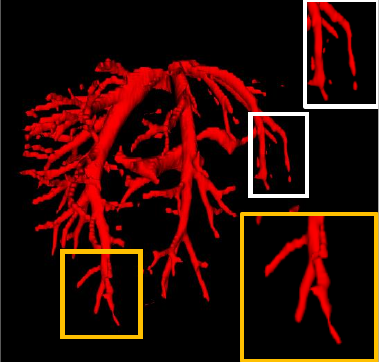}&
    \includegraphics[width=0.1\textwidth]{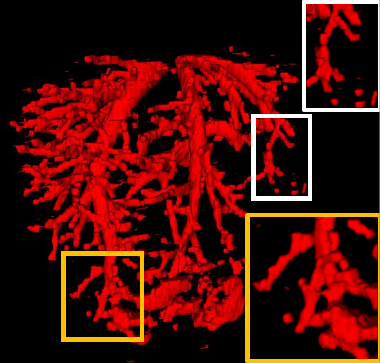} &
    \includegraphics[width=0.1\textwidth]{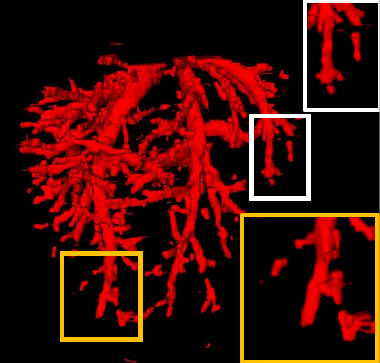} &
    \includegraphics[width=0.1\textwidth]{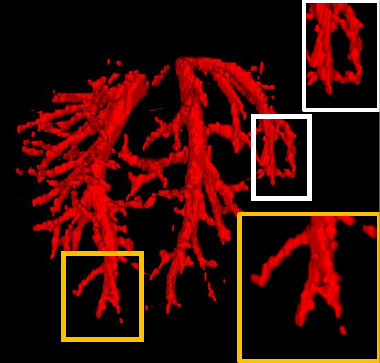} &
    \includegraphics[width=0.1\textwidth]{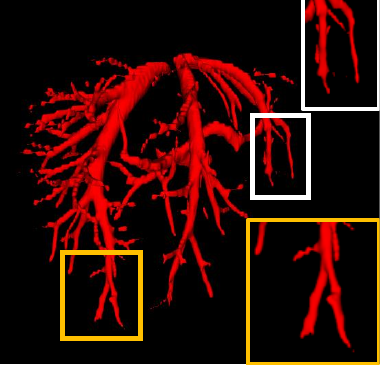} &
    \includegraphics[width=0.1\textwidth]{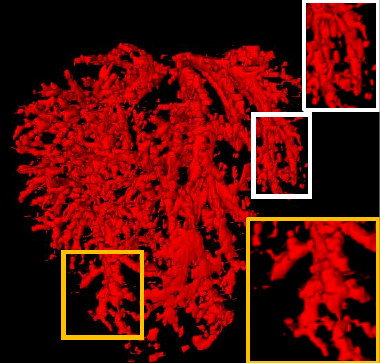} &
    \includegraphics[width=0.1\textwidth]{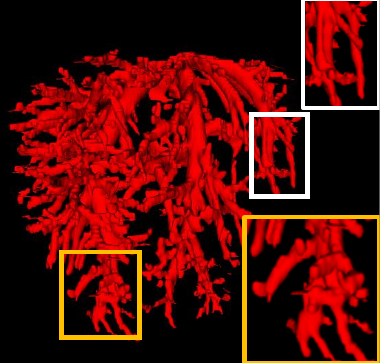} \\
    \includegraphics[width=0.1\textwidth]{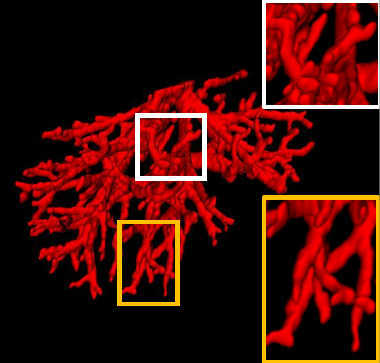} &
    \includegraphics[width=0.1\textwidth]{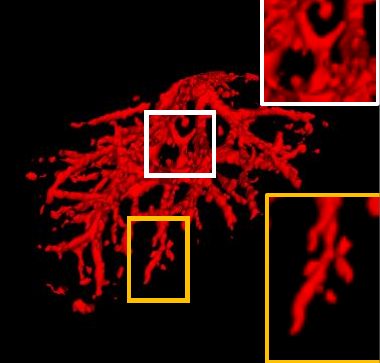}&
    \includegraphics[width=0.1\textwidth]{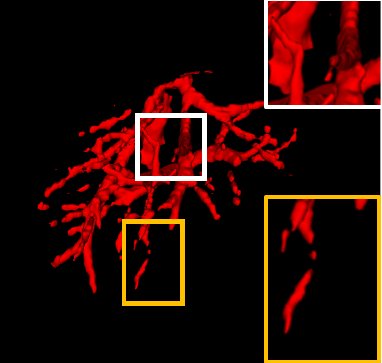}&
    \includegraphics[width=0.1\textwidth]{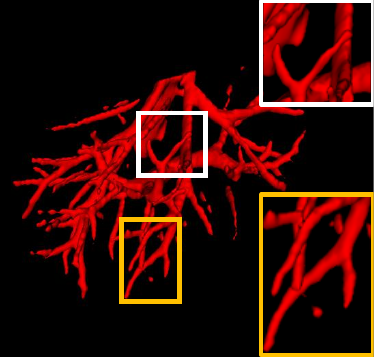}&
    \includegraphics[width=0.1\textwidth]{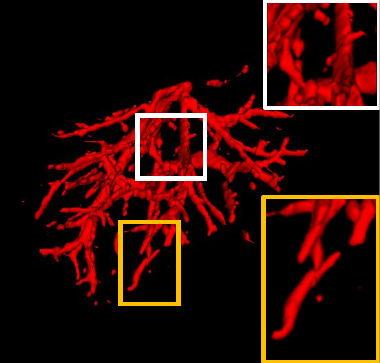} &
    \includegraphics[width=0.1\textwidth]{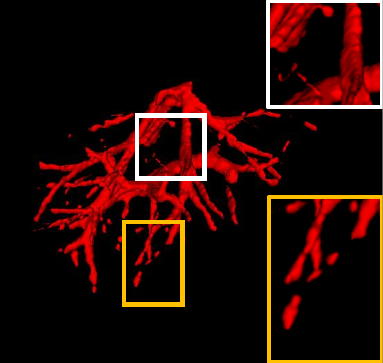} &
    \includegraphics[width=0.1\textwidth]{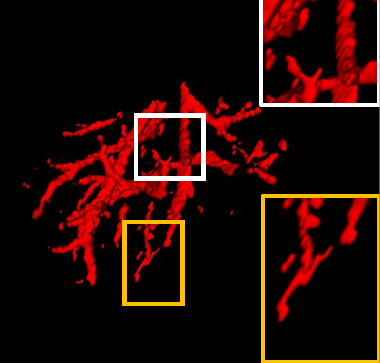} &
    \includegraphics[width=0.1\textwidth]{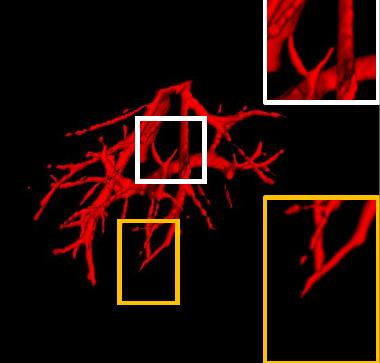} &
    \includegraphics[width=0.1\textwidth]{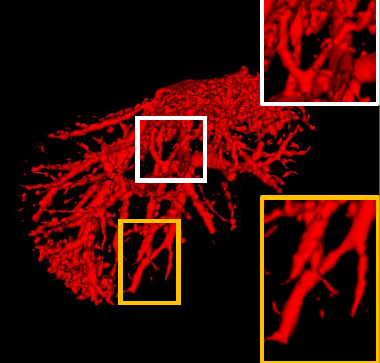} &
    \includegraphics[width=0.1\textwidth]{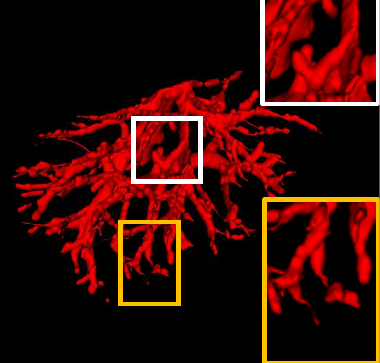} \\
    \includegraphics[width=0.1\textwidth]{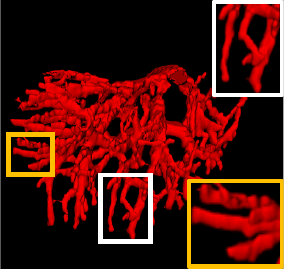} &
    \includegraphics[width=0.1\textwidth]{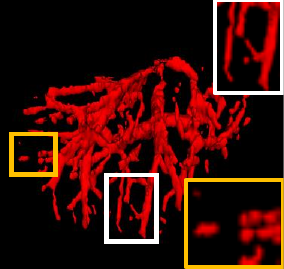}&
    \includegraphics[width=0.1\textwidth]{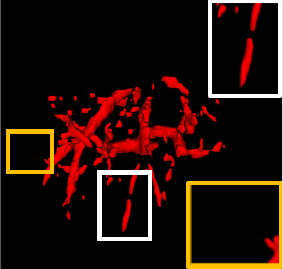}&
    \includegraphics[width=0.1\textwidth]{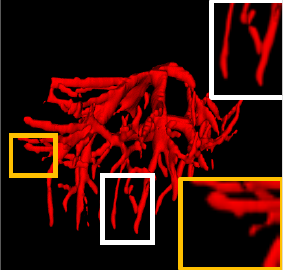}&
    \includegraphics[width=0.1\textwidth]{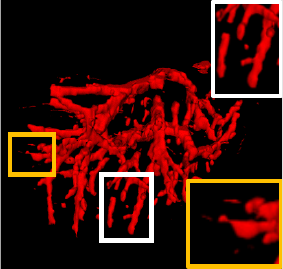} &
    \includegraphics[width=0.1\textwidth]{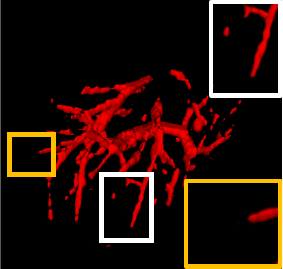} &
    \includegraphics[width=0.1\textwidth]{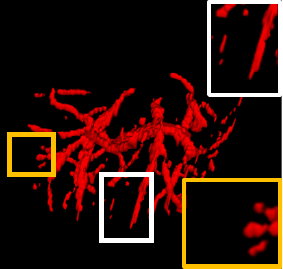} &
    \includegraphics[width=0.1\textwidth]{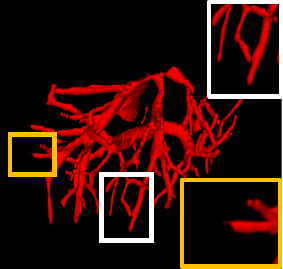} &
    \includegraphics[width=0.1\textwidth]{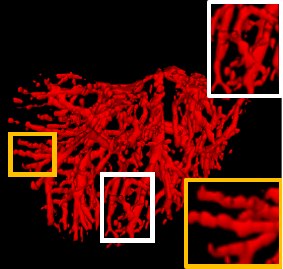} &
    \includegraphics[width=0.1\textwidth]{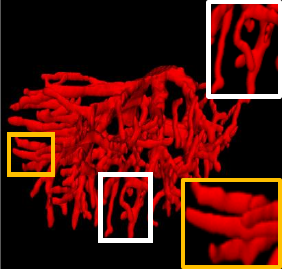} \\
   \scriptsize{(a) } & 
   \scriptsize{\update{(b)} } & 
   \scriptsize{\update{(c)} } & 
   \scriptsize{\updateSecond{(d)} } & 
   \scriptsize{(e) } & 
   \scriptsize{(f) } & 
   \scriptsize{(g) } & 
   \scriptsize{(h) } &
   \scriptsize{(i) } &
   \scriptsize{(j) }\\
   \tiny{Ours} & 
   \tiny{\update{\textsl{HiDiff}}} &
   \tiny{\update{\textsl{MERIT}}} & 
   \tiny{\updateSecond{\textsl{TransUNet}}} & 
   \tiny{\textsl{MedSegDiff}} & 
   \tiny{\textsl{EnsemDiff}} & 
   \tiny{\textsl{SwinUNETR}} & 
   \tiny{\textsl{nnUNet} } & 
   \tiny{\textsl{Gao et al.} } & 
   \tiny{Ground truth}
  \end{tabular}
  \caption{
  \textbf{Visualizations on the \textsl{3D-ircadb-01}\cite{b40} dataset.}
  (a) The liver vessel tree segmented by our proposed model; (b), (c), (d), (e), (f), (g), (h) and (i) are the liver vessel tree segmented by the baselines: \update{\textsl{HiDiff}}\cite{b63}, \update{\textsl{MERIT}}\cite{b62}, \updateSecond{\textsl{TransUNet}}\cite{b68}, \textsl{MedSegDiff}\cite{b20}, \textsl{EnsemDiff}\cite{b19}, \textsl{Swin UNETR}\cite{b28}, \textsl{nnUNet}\cite{b27} and \cite{b41} respectively; 
  (j) The ground truth liver vessel tree.
  The yellow and white boxes compare the completeness and continuity between our proposed model and the baselines. 
  We also show the boxes enlarged in the top\slash bottom right corners. 
  Our method (1$st$ column) achieves the most similar predictions to the ground truth (last column) in both fine vessel segmentation, and connectivity.
  }
  \label{fig:resfig1}
\end{figure*}

\medskip\noindent\textbf{Qualitative evaluation on \textsl{3D-ircadb-01}.}
In \fig{resfig1} we provide a qualitative evaluation on three test cases from the \textsl{3D-ircadb-01} dataset.
Compared to the other methods, the appearance of our predicted vessel-tree segmentation (first column) is the most similar to the ground truth (last column). 
The vascular structures marked by the yellow and white boxes in \fig{resfig1} are almost completely segmented by our method, while the other methods oversegment or miss parts of the vessel tree.
Especially the segmentation results of \update{\textsl{MERIT} \cite{b62}}, \textsl{Swin UNETR} \cite{b28} and \textsl{nnUNet} \cite{b27}, are visibly sparse. 
These results relate to the low \textsl{DSC} and \textsl{Sen} scores of \update{\textsl{MERIT}}, \textsl{Swin UNETR} and \textsl{nnUNet} in \tab{results1}. 
Although the vessel structures of \textsl{EnsemDiff} \cite{b19} (sixth column) and \textsl{MedSegDiff} \cite{b20} (fifth column) are denser than those of the non-diffusion methods, they provide more discontinuous vessel masks at the extremities of the tree (\ie for smaller vessels).

In \fig{resfig1} we also highlight the connectivity of the distal liver vessel branches in the yellow boxes. 
The visualizations show that our segmentation is as continuous as the ground truth.
This corresponds to a \textsl{Con} score closer to $1$ in \tab{results1}. 
Comparing the vessel branches of \textsl{nnUNet} with the other methods for the distal vessels, we see that these are precise, being exceeded only by our method. 
This is again consistent with the \textsl{Con} score of \textsl{nnUNet} in \tab{results1} -- the second best score.

\begin{figure*}[t]
     \begin{tabular}{ll}    
        \resizebox{.7\linewidth}{!}{ 
            \begin{booktabs}{l llllll}
                \toprule
                & Repr. & DSC (\%) & Sen (\%) & Spe (\%) & \updateSecond{$sec.$} & \updateSecond{steps} \\ 
                & type  &           &          &     &\updateSecond{$/slice$} & \\ 
                \midrule
                Gao et al.\cite{b41} &{2D} &{$73.14\pm11.44$} &{$76.84\pm11.57$} &{$99.63\pm0.34$} & \updateSecond{$<0.1$} & - \\
                \midrule
                \textsl{nnUNet}\cite{b27} &{3D} &{$81.39\pm5.04$} &{$76.06\pm7.48$} &{$99.91\pm0.05$} & \updateSecond{$<0.1$} & -\\
                \textsl{Swin UNETR}\cite{b28} &{3D} &{$65.54\pm6.65$} &{$62.71\pm10.31$} &{$99.76\pm0.13$} & \updateSecond{$<0.1$} & -\\
                \updateSecond{\textsl{TransUNet}}\cite{b68} &\updateSecond{3D} &
                \updateSecond{$78.21\pm6.17$} &\updateSecond{$75.11\pm7.58$} &
                \updateSecond{$99.86\pm0.08$} & \updateSecond{$<0.1$} & -\\
                \midrule
                \textsl{EnsemDiff}\cite{b19} & {2.5D} & {$70.00\pm9.20$} & {$56.27\pm10.81$}  & {$\textbf{99.97}\pm0.03$} & \updateSecond{$20.67$} & \updateSecond{$1000$} \\
                \textsl{MedSegDiff}\cite{b20}  & {2.5D} & {$76.85\pm7.03$} & {$65.96\pm9.43$} & {$99.96\pm0.04$} & \updateSecond{$26.92$} & \updateSecond{$1000$}\\ 
                \update{\textsl{MERIT}}\cite{b62} & \update{2.5D} &\update{$69.54\pm6.72$} &\update{$56.63\pm8.43$} &\update{$99.95\pm0.04$} & \updateSecond{$<0.1$} & -\\
                \update{\textsl{HiDiff}}\cite{b63} & \update{2.5D} &\update{$70.53\pm6.66$} &\update{$78.32\pm5.49$} &\update{$99.63\pm0.12$} & \updateSecond{$0.25$} & \updateSecond{$10$}\\
                \midrule[dashed]
                 Ours & {2.5D} & {$\textbf{81.41}\pm6.64$} & {$\textbf{81.35}\pm6.93$} & {$99.84\pm0.12$} & \updateSecond{$43.21$} & \updateSecond{$1000$}\\ 
                \bottomrule
            \end{booktabs}
        } &
        \begin{tabular}{l}
        \hspace*{-15px}
        \includegraphics[width=.3\linewidth]{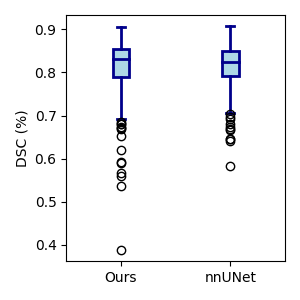} 
        \end{tabular}
        \\
    \end{tabular}    
    \caption{%
        \textbf{Results on the LiVS \cite{b41} dataset.}  
        \emph{Left:} The performance of our method compared with \update{\textsl{HiDiff}}\cite{b63}, \update{\textsl{MERIT}}\cite{b62}, \updateSecond{\textsl{TransUNet}}\cite{b68}, \textsl{MedSegDiff}\cite{b20}, \textsl{EnsemDiff}\cite{b19}, \textsl{Swin UNETR}\cite{b28}, \textsl{nnUNet}\cite{b27} and Gao et al.\cite{b41}.  
        \textsl{clDice} and \textsl{Con} are not applicable for the \textsl{LiVS} dataset with discontinuous annotations.
        Our method outperforms others in terms of \textsl{Sen} scores.  
        \textsl{Spe} of our model is slightly lower than \textsl{EnsemDiff}, which is a trade-off between completeness (\textsl{Sen}) and accuracy (\textsl{Spe}) of vessel segmentation.
        \emph{Right:} Our \textsl{DSC} score is comparable with \textsl{nnUNet} is because that the high-contrast tumors in the liver will cause more outliers (shown in the box plot) for our method, and negatively affect the averaged \textsl{DSC} score over all data.
        \updateSecond{Deterministic methods are more efficient than generative methods in inference. }
    }
    \label{fig:num_LiVS}
\end{figure*}
\begin{figure*}[t]
  \centering
  \begin{tabular}{c@{\hskip 0.01in}c@{\hskip 0.01in}c@{\hskip 0.01in}c@{\hskip 0.01in}c@{\hskip 0.01in}c@{\hskip 0.01in}c@{\hskip 0.01in}c@{\hskip 0.01in}c@{\hskip 0.01in}c}
    \includegraphics[width=0.1\textwidth]{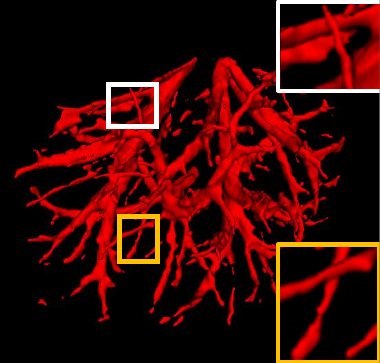} &
    \includegraphics[width=0.1\textwidth]{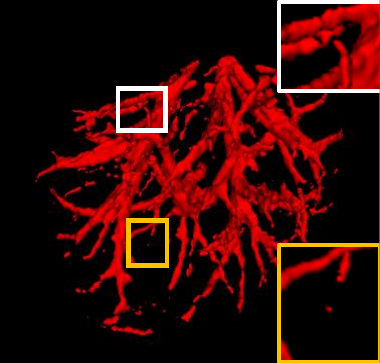}&
    \includegraphics[width=0.1\textwidth]{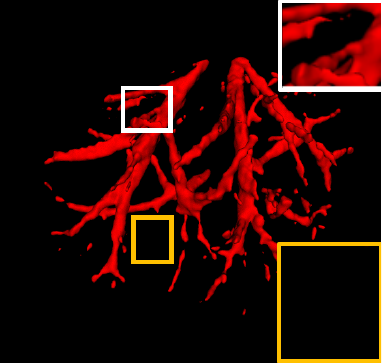}&
    \includegraphics[width=0.1\textwidth]{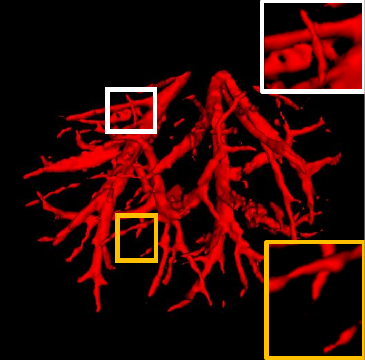}&
    \includegraphics[width=0.1\textwidth]{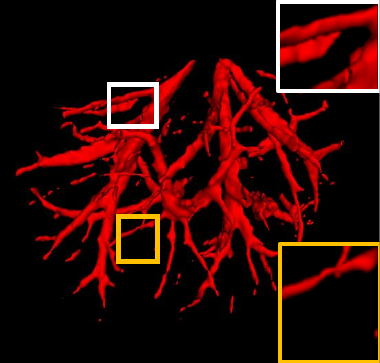} &
    \includegraphics[width=0.1\textwidth]{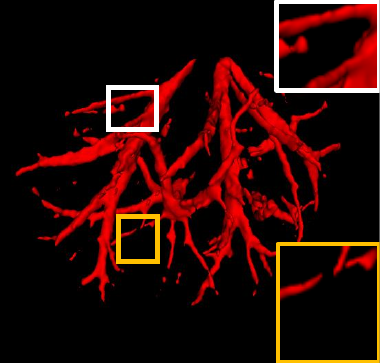} &
    \includegraphics[width=0.1\textwidth]{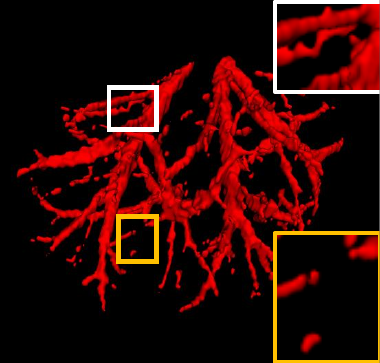} &
    \includegraphics[width=0.1\textwidth]{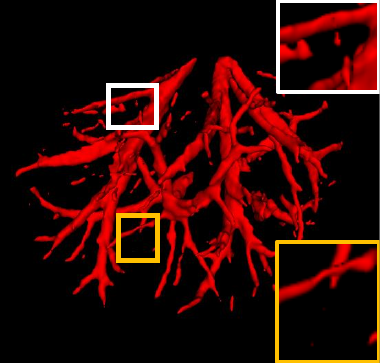} &
    \includegraphics[width=0.1\textwidth]{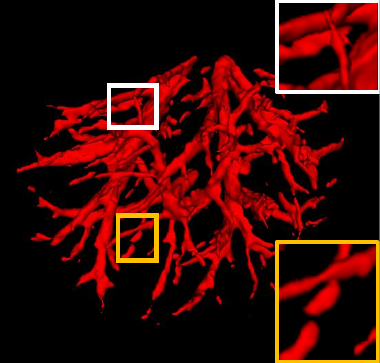} &
    \includegraphics[width=0.1\textwidth]{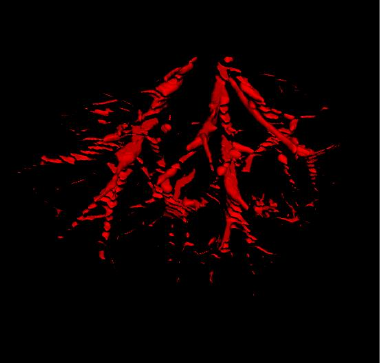} \\
    \includegraphics[width=0.1\textwidth]{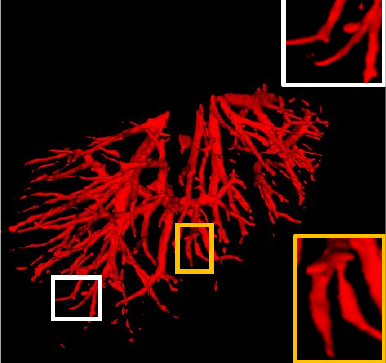} &
    \includegraphics[width=0.1\textwidth]{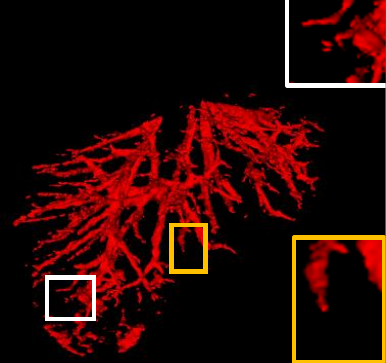}&
    \includegraphics[width=0.1\textwidth]{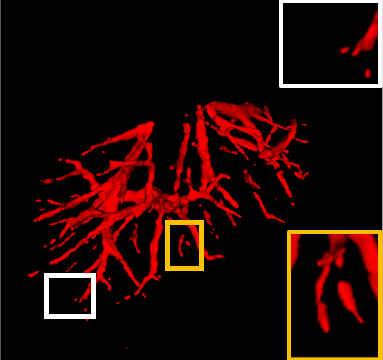}&
    \includegraphics[width=0.1\textwidth]{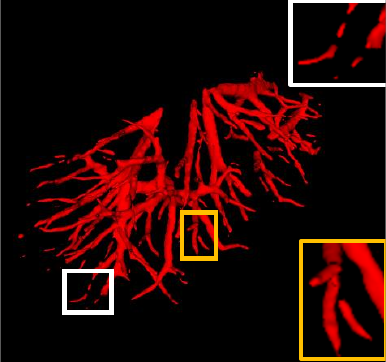}&
    \includegraphics[width=0.1\textwidth]{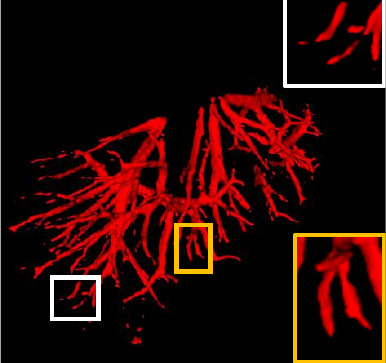} &
    \includegraphics[width=0.1\textwidth]{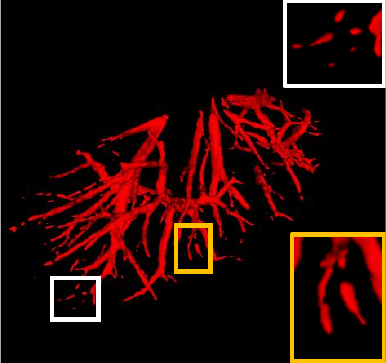} &
    \includegraphics[width=0.1\textwidth]{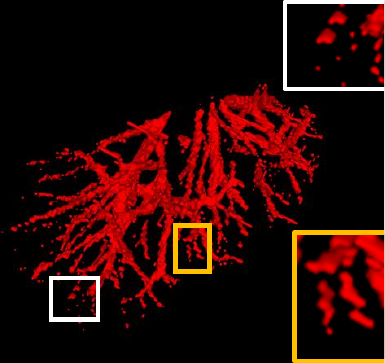} &
    \includegraphics[width=0.1\textwidth]{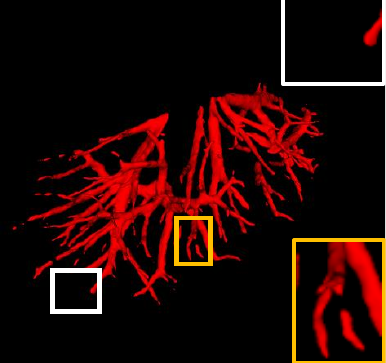} &
    \includegraphics[width=0.1\textwidth]{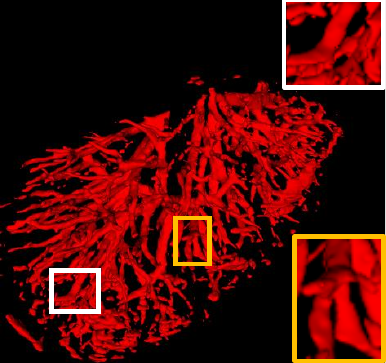} &
    \includegraphics[width=0.1\textwidth]{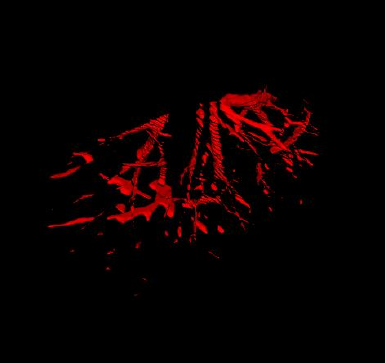} \\
    \includegraphics[width=0.1\textwidth]{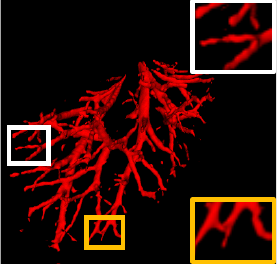} &
    \includegraphics[width=0.1\textwidth]{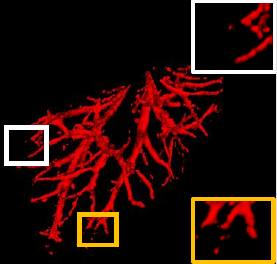}&
    \includegraphics[width=0.1\textwidth]{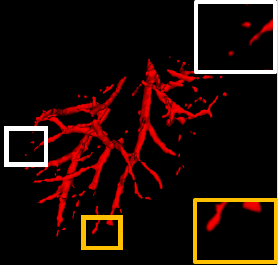}&
    \includegraphics[width=0.1\textwidth]{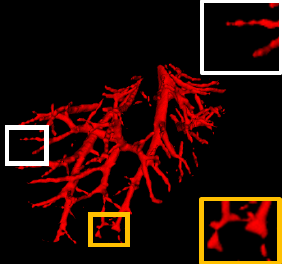}&
    \includegraphics[width=0.1\textwidth]{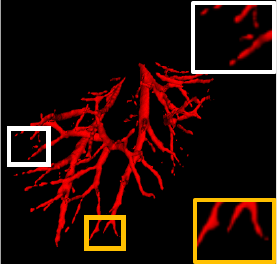} &
    \includegraphics[width=0.1\textwidth]{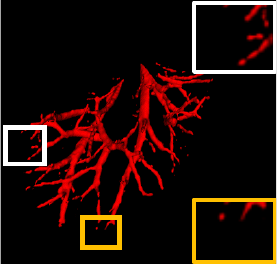} &
    \includegraphics[width=0.1\textwidth]{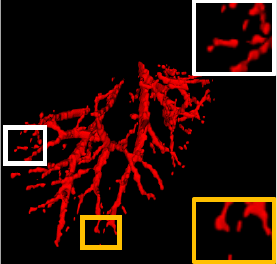} &
    \includegraphics[width=0.1\textwidth]{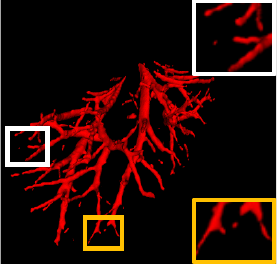} &
    \includegraphics[width=0.1\textwidth]{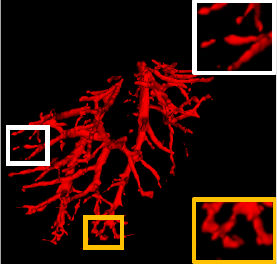} &
    \includegraphics[width=0.1\textwidth]{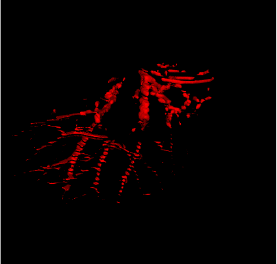} \\
   \scriptsize{(a) } & 
   \scriptsize{\update{(b)} } & 
   \scriptsize{\update{(c)} } & 
   \scriptsize{\updateSecond{(d)} } &
   \scriptsize{(e) } & 
   \scriptsize{(f) } & 
   \scriptsize{(g) } & 
   \scriptsize{(h) } &
   \scriptsize{(i) } &
   \scriptsize{(j) } \\
   \tiny{Ours} & 
   \tiny{\update{\textsl{HiDiff}}} &
   \tiny{\update{\textsl{MERIT}}} &
   \tiny{\updateSecond{\textsl{TransUNet}}} &
   \tiny{\textsl{MedSegDiff}} & 
   \tiny{\textsl{EnsemDiff}} & 
   \tiny{\textsl{SwinUNETR}} & 
   \tiny{\textsl{nnUNet} } & 
   \tiny{\textsl{Gao et al.} } & 
   \tiny{Ground truth}
  \end{tabular}
    \caption{
        \textbf{Visualizations of the \textsl{LiVS} \cite{b41} dataset}. 
        (a) The liver vessel tree segmented by our proposed model; 
        (b)--(i) are the liver vessel tree segmented by the baselines: 
        \update{\textsl{HiDiff}} \cite{b63}, \update{\textsl{MERIT}} \cite{b62}, \updateSecond{\textsl{TransUNet}} \cite{b68}, \textsl{MedSegDiff} \cite{b20}, \textsl{EnsemDiff} \cite{b19}, \textsl{Swin UNETR} \cite{b28}, \textsl{nnUNet} \cite{b27} and Gao et al. \cite{b41};
        (j) The discontinuous (partially annotated) ground truth liver vessel tree.
        Yellow and white boxes show the completeness and continuity between our proposed model and baselines.
        These boxes are enlarged in the top/bottom right corner.
        }
  \label{fig:5}
\end{figure*}
\begin{figure*}[h!]
    \centering
      \begin{tabular}{c@{\hskip 0.01in}c@{\hskip 0.01in}c@{\hskip 0.01in}c@{\hskip 0.01in}c@{\hskip 0.01in}c@{\hskip 0.01in}c@{\hskip 0.01in}c@{\hskip 0.01in}c@{\hskip 0.01in}c@{\hskip 0.01in}c}
    \includegraphics[width=0.1\textwidth]{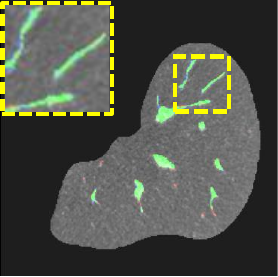} &
    \includegraphics[width=0.1\textwidth]{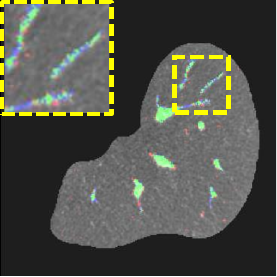} & 
    \includegraphics[width=0.1\textwidth]{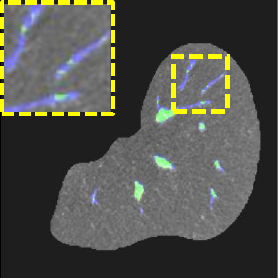} &     
    \includegraphics[width=0.1\textwidth]{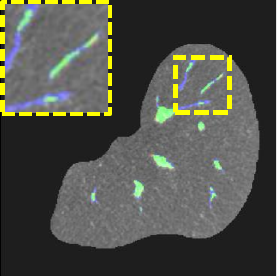} &   
    \includegraphics[width=0.1\textwidth]{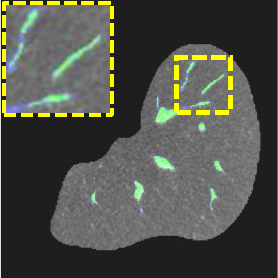} & 
    \includegraphics[width=0.1\textwidth]{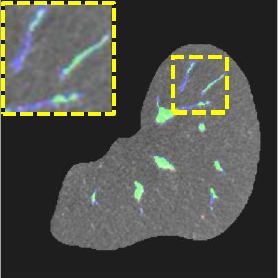} & 
    \includegraphics[width=0.1\textwidth]{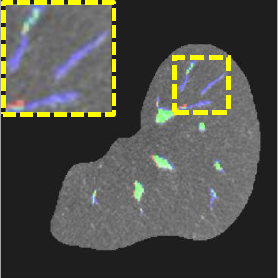} &
    \includegraphics[width=0.1\textwidth]{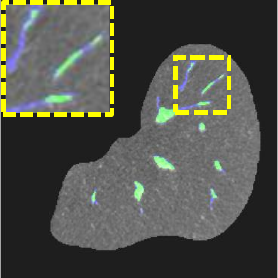} &
    \includegraphics[width=0.1\textwidth]{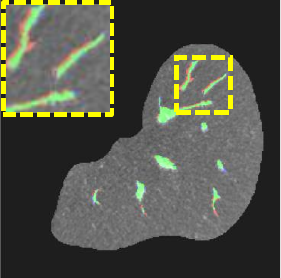} &
    \includegraphics[width=0.1\textwidth]{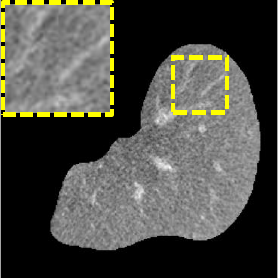} \\
    \includegraphics[width=0.1\textwidth]{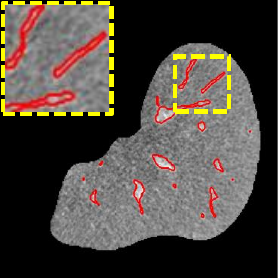} &
    \includegraphics[width=0.1\textwidth]{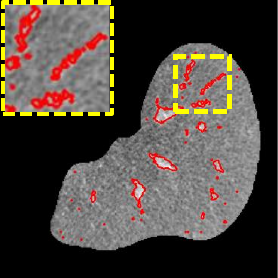} & 
    \includegraphics[width=0.1\textwidth]{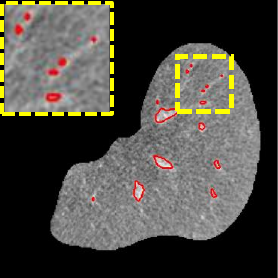} & 
    \includegraphics[width=0.1\textwidth]{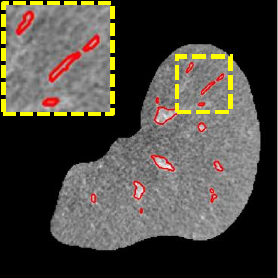} & 
    \includegraphics[width=0.1\textwidth]{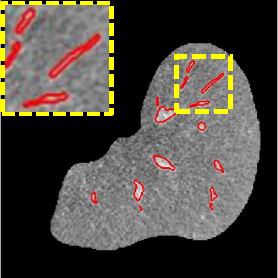} & 
    \includegraphics[width=0.1\textwidth]{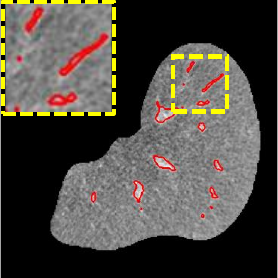} & 
    \includegraphics[width=0.1\textwidth]{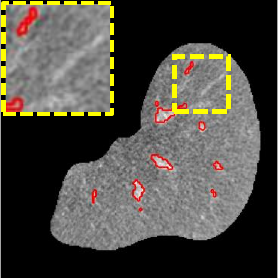} &
    \includegraphics[width=0.1\textwidth]{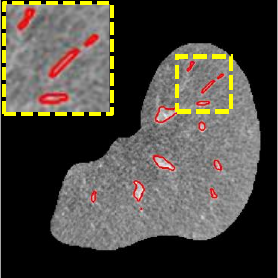} &
    \includegraphics[width=0.1\textwidth]{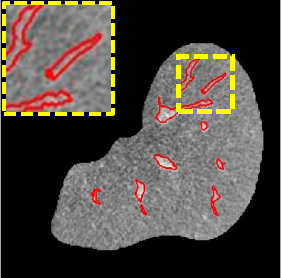} &
    \includegraphics[width=0.1\textwidth]{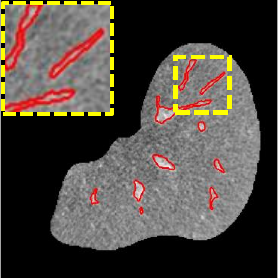} \\
    \includegraphics[width=0.1\textwidth]{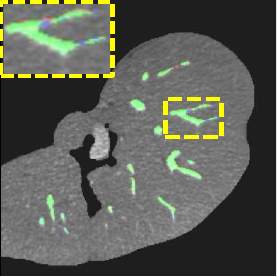} &
    \includegraphics[width=0.1\textwidth]{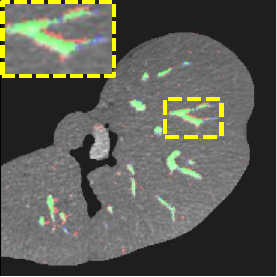} & 
    \includegraphics[width=0.1\textwidth]{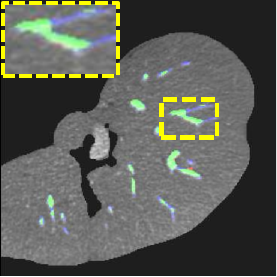} & 
    \includegraphics[width=0.1\textwidth]{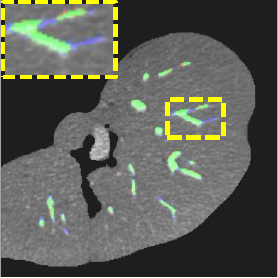} & 
    \includegraphics[width=0.1\textwidth]{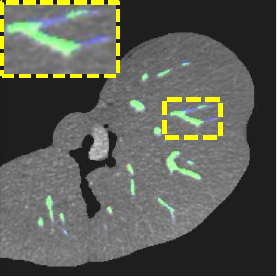} & 
    \includegraphics[width=0.1\textwidth]{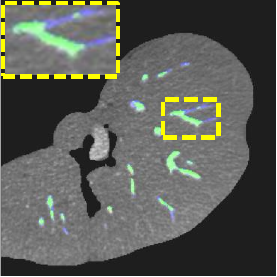} & 
    \includegraphics[width=0.1\textwidth]{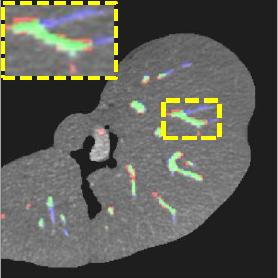} &
    \includegraphics[width=0.1\textwidth]{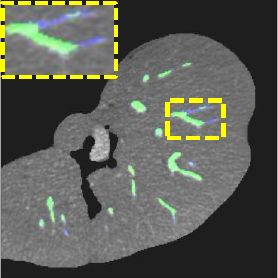} &
    \includegraphics[width=0.1\textwidth]{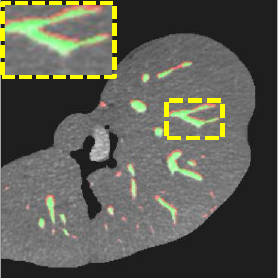} &
    \includegraphics[width=0.1\textwidth]{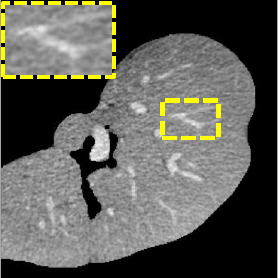} \\
    \includegraphics[width=0.1\textwidth]{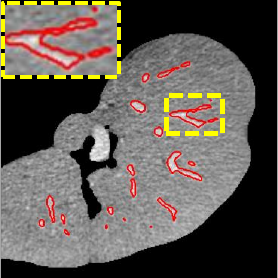} &
    \includegraphics[width=0.1\textwidth]{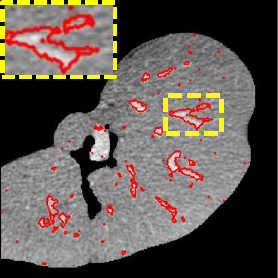} & 
    \includegraphics[width=0.1\textwidth]{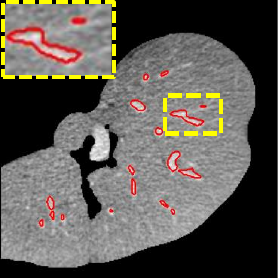} & 
    \includegraphics[width=0.1\textwidth]{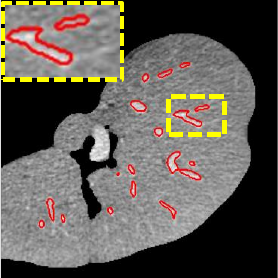} & 
    \includegraphics[width=0.1\textwidth]{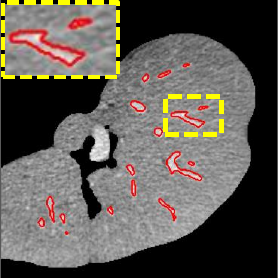} & 
    \includegraphics[width=0.1\textwidth]{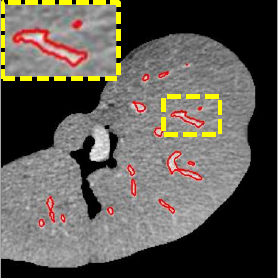} & 
    \includegraphics[width=0.1\textwidth]{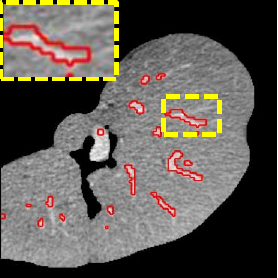} &
    \includegraphics[width=0.1\textwidth]{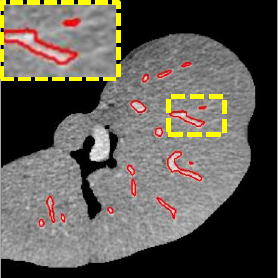} &
    \includegraphics[width=0.1\textwidth]{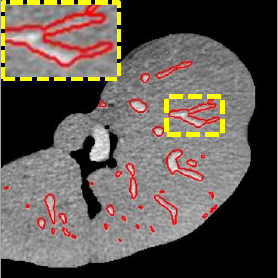} &
    \includegraphics[width=0.1\textwidth]{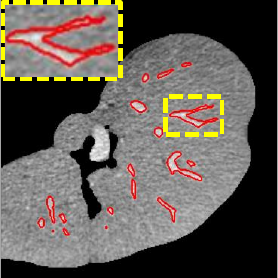} \\
    \includegraphics[width=0.1\textwidth]{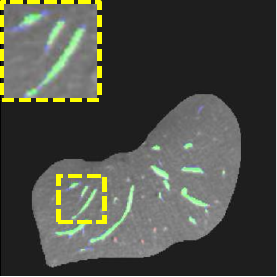} &
    \includegraphics[width=0.1\textwidth]{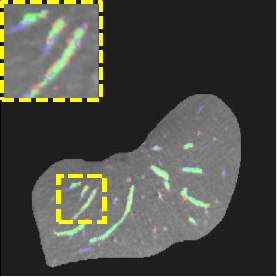} & 
    \includegraphics[width=0.1\textwidth]{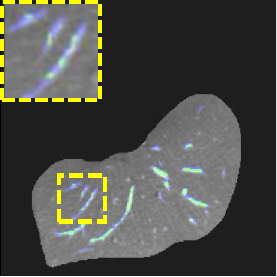} & 
    \includegraphics[width=0.1\textwidth]{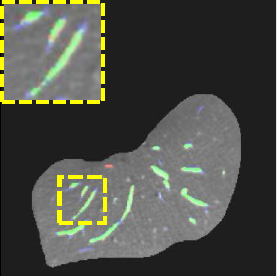} & 
    \includegraphics[width=0.1\textwidth]{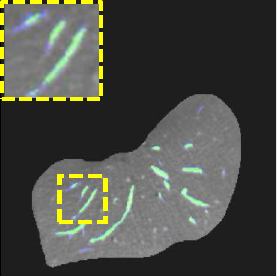} & 
    \includegraphics[width=0.1\textwidth]{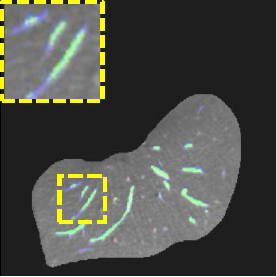} & 
    \includegraphics[width=0.1\textwidth]{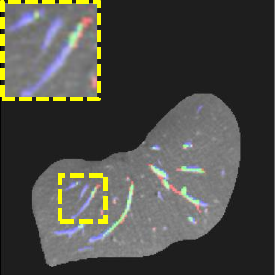} &
    \includegraphics[width=0.1\textwidth]{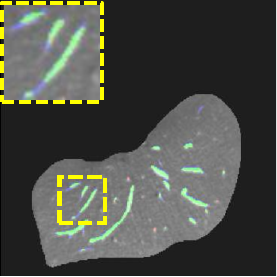} &
    \includegraphics[width=0.1\textwidth]{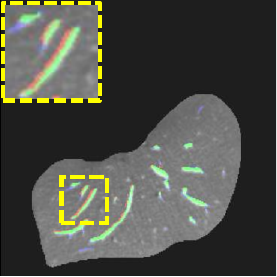} &
    \includegraphics[width=0.1\textwidth]{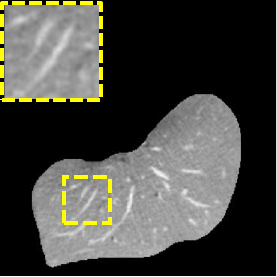} \\
    \includegraphics[width=0.1\textwidth]{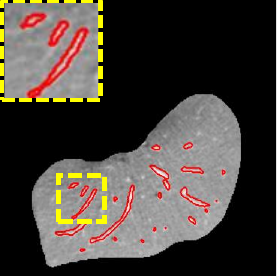} &
    \includegraphics[width=0.1\textwidth]{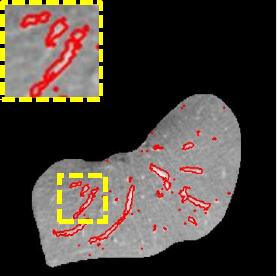} & 
    \includegraphics[width=0.1\textwidth]{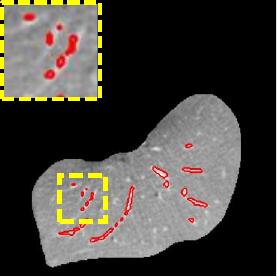} & 
    \includegraphics[width=0.1\textwidth]{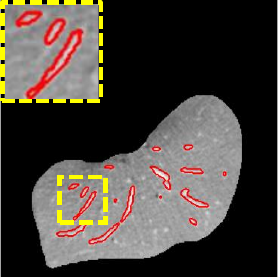} & 
    \includegraphics[width=0.1\textwidth]{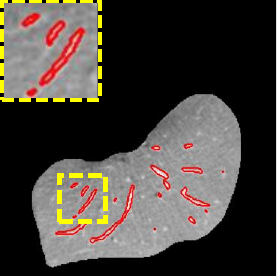} & 
    \includegraphics[width=0.1\textwidth]{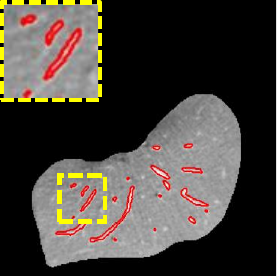} & 
    \includegraphics[width=0.1\textwidth]{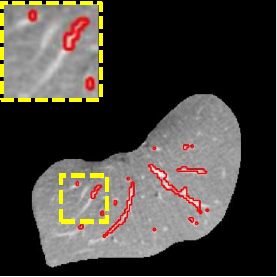} &
    \includegraphics[width=0.1\textwidth]{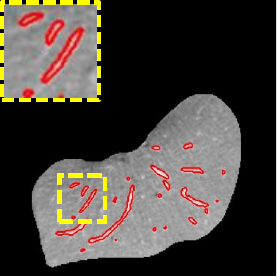} &
    \includegraphics[width=0.1\textwidth]{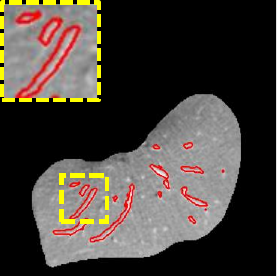} &
    \includegraphics[width=0.1\textwidth]{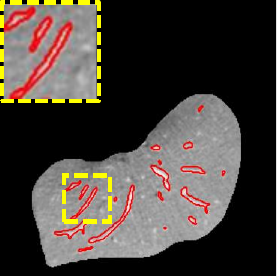} \\
    \scriptsize{(a)} &
    \scriptsize{\update{(b)}} &
    \scriptsize{\update{(c)}} &
    \scriptsize{\updateSecond{(d)}} &
    \scriptsize{(e)} &
    \scriptsize{(f)} &
    \scriptsize{(g)} &
    \scriptsize{(h)} &
    \scriptsize{(i)} &
    \scriptsize{(j)}\\
    \tiny{Ours} &
    \tiny{\update{\textsl{HiDiff}}} &
    \tiny{\update{\textsl{MERIT}}} &
    \tiny{\updateSecond{\textsl{TransUNet}}} &
    \tiny{\textsl{MedSegDiff}} &
    \tiny{\textsl{EnsemDiff}} &
    \tiny{\textsl{SwinUNetr}} &
    \tiny{\textsl{nnUNet}} &
    \tiny{\textsl{Gao et al.}} &
    \tiny{\textsl{Ground truth} } \\
    
    \end{tabular}    
    \caption{
        \textbf{Cross-sectional visualizations on \textsl{LiVS} \cite{b41} dataset}. 
        We compare segmentation masks from our proposed model with: 
        \update{\textsl{HiDiff}} \cite{b63}, \update{\textsl{MERIT}} \cite{b62}, \updateSecond{\textsl{TransUNet}} \cite{b68}, \textsl{MedSegDiff} \cite{b20}, \textsl{EnsemDiff} \cite{b19}, \textsl{Swin UNETR} \cite{b28}, \textsl{nnUNet} \cite{b27} and Gao et al. \cite{b41}.
        The first, third and fifth rows are the overlaid predicted segmentation masks on top of a CT slice (green: true positive, red: false positive, blue: false negative);
        The second, fourth and final rows are the contours obtained from the predictions of different methods, when compared to the ground truth.
        We highlight in the yellow box differences in predictions.
        Our method can segment fine vessel structures and preserve their connectivity, while baselines fail in these situations.
        }
  \label{fig:cross_LiVS}
\end{figure*}

\subsection{Comparative experiments on \textsl{LiVS} dataset}
\noindent\textbf{Quantitative evaluation on \textsl{LiVS}.}
\fig{num_LiVS} (left) compares our method with \update{\textsl{HiDiff} \cite{b63}}, \update{\textsl{MERIT} \cite{b62}}, \textsl{MedSegDiff} \cite{b20}, \updateSecond{\textsl{TransUNet}} \cite{b68}, \textsl{EnsemDiff} \cite{b19}, \textsl{Swin UNETR} \cite{b28}, \textsl{nnUNet} \cite{b27} and Gao et al. \cite{b41}. 
The ground truth of the liver vessels is discontinuous in the longitudinal direction, as shown in \fig{5}(j). 
On \textsl{LiVS} we cannot report \textsl{clDice} and \textsl{Con} scores, because of the discontinuous annotations.
In \fig{num_LiVS}, our method has the best \textsl{DSC} and \textsl{Sen} scores. 
Especially in terms of \textsl{Sen}, our method performs better than the baselines, indicating a more complete vessel segmentation. 
In terms of \textsl{DSC} scores, we are comparable to \textsl{nnUNet} because the \textsl{DSC} scores are susceptible to outliers, as shown in \fig{num_LiVS} (right).
The methods of \update{\textsl{HiDiff} and} Gao \etal have comparable \textsl{Sen} scores to \textsl{nnUNet}, but their \textsl{DSC} score are lower, because of having the lowest \textsl{Spe} score of all methods.
\textsl{SwinUNetr} has the lowest \textsl{DSC} scores, while \update{\textsl{MERIT} and \textsl{EnsemDiff} have the lowest} \textsl{Sen} scores.
As with the \textsl{3D-ircadb-01} dataset, our improvements are due to the graph-attention conditioning. 
Although \update{\textsl{HiDiff} and} \textsl{MedSegDiff} have lower scores than our model in terms of \textsl{DSC} and \textsl{Sen}, they outperform the \textsl{EnsemDiff} model. 
This may be due to the dynamic conditioning method used in \update{\textsl{HiDiff} and} \textsl{MedSegDiff}.
The standard deviations of our predictions and \update{predictions of \textsl{HiDiff} and} \textsl{MedSegDiff} are comparable, and they are both lower than the standard deviation of \textsl{EnsemDiff}.
This could indicate that the dynamic conditioning diffusion model (ours, \update{\textsl{HiDiff}} and \textsl{MedSegDiff}) can provide more stable results than the vanilla conditioning model (\textsl{EnsemDiff}). 
\updateSecond{However, these diffusion-based methods are less efficient than deterministic methods during inference, as shown in the \fig{num_LiVS}.
Our model is slower than \textsl{MedSegDiff} and \textsl{EnsemDiff} when the inference batch size is set to 50 slices.
This slowdown is due to the additional cost of graph attention and local feature integration (LFI) in our model.
The graph attention introduces anatomic structure into the model, improving accuracy, but also increases inference time compared to other diffusion baselines.
Similarly, LFI compensates for the sparse node representation, but the grid sample layer used in LFI results in heavier computation.}

\medskip\noindent\textbf{Qualitative evaluation on \textsl{LiVS}.}
In \fig{5}, we provide a qualitative comparison in $3$D on three test cases between our model and the eight baselines. 
Although it is hard to directly compare the segmentations of the different methods, due to the discontinuous ground truth (shown in \fig{5}(j)), inter-method comparison can still be informative.
The yellow and white boxes in \fig{5} show that most baselines predict fractured small vessels for the distal branch, while our segmented vessel structures are denser and more complete. 

We compare vessel completeness in \fig{5}, where we show a qualitative comparison for the slices of the three cases in \fig{cross_LiVS}, in a $2$D cross-sectional view. 
The first, third and fifth rows show the overlayed segmentation masks of the liver vessel mask of different methods, when compared to the ground truth mask. 
Green, red and blue colors represent true positive, false positive and false negative, respectively.
The blue areas of the overlaid masks from columns (b) to (i) reflect the missing contours of the baselines. 
Our model has fewer blue regions, indicating that our segmentation is more complete for both the large and small vessel structures. 
The red contours in the second, forth and final rows of \fig{cross_LiVS} outline the boundary of the predicted liver vessel segmentation of different methods. 
The contours in the last column (j) correspond to the boundary of ground truth vessel mask. 
We enlarged the regions using the yellow boxes, to highlight differences in the vessel completeness.
Comparing the contours of the baselines from column (b) to (i) with the ground truth, tiny vessel blobs are not outlined. 
For the baseline (i), the contours reflect a heavy oversegmentation.
Our model benefits from the long-range feature dependency encoded in the multi-scale graph-attention conditioning.

\subsection{Model ablation study}
\label{sec:abl}
\begin{table*}[b]
    \caption{\textbf{Ablation study on \textsl{3D-ircadb-01} dataset.}
        We ablate the effect of each individual component presented in \fig{detailed} and the effect of the post-processing:
        (A) vanilla conditioning model \cite{b19};
        (B) dynamic conditioning model; 
        (C) multiscale graph-attention conditioning;
        and (D) post-processing.
        Interestingly, the \textsl{(A) vanilla model} and the \textsl{(A) vanilla model} combined with \textsl{(B) dynamic conditioning} perform best in terms of \textsl{Spe} scores.
        This is because predicting less structures (low \textsl{Sen} scores) entails fewer false positives. 
        However, the overall combination performs best in \textsl{DSC} score.
    }
\centering 
    \resizebox{1\linewidth}{!}{
    \begin{tabular}{ccccccc}
    \toprule
    (A) Vanilla  & (B) Dynamic   & (C) Dynamic multiscale &(D) Post-& \multirow{2}{*}{DSC(\%)} & \multirow{2}{*}{Sen(\%)} & \multirow{2}{*}{Spe(\%)} \\  
    conditioning & conditioning   & graph-attention    & processing   
    &      &    & \\
    \cmidrule(lr){1-4}\cmidrule(lr){5-7}
    \checkmark& $\times$& $\times$& $\times$
    & $55.07\pm9.70$& $40.45\pm10.6$& $99.98\pm0.02$\\
    \checkmark& \checkmark& $\times$& $\times$
    & $59.00\pm5.43$& $44.22\pm6.19$& $\textbf{99.98}\pm0.01$\\
    \checkmark& \checkmark& \checkmark& $\times$
    &$62.69\pm3.93$ &$\textbf{79.80}\pm6.33$ &$99.70\pm0.17$ \\
    \checkmark& \checkmark& \checkmark& \checkmark
    & $\textbf{71.26}\pm1.93$& $71.59\pm4.07$& $99.89\pm0.04$ \\
    \bottomrule
\end{tabular}}
\label{tab:4}
\end{table*}
\begin{table*}[t]
    \caption{\textbf{Ablation study on \textsl{LiVS} dataset.}
        We ablate the effect of each individual component presented in \fig{detailed} and the effect of the post-processing:
        (A) vanilla conditioning model \cite{b19};
        (B) dynamic conditioning model; 
        (C) multiscale graph-attention conditioning;
        and (D) post-processing.
        Inference ensembling is used due to the discontinuous annotations of the dataset.
        Post-processing is not mandatory for this kind of dataset. 
    }
\centering 
    \resizebox{1\linewidth}{!}{
    \begin{tabular}{ccccccc}
    \toprule
    (A) Vanilla  & (B) Dynamic   & (C) Dynamic multiscale &(D) Post-& \multirow{2}{*}{DSC(\%)} & \multirow{2}{*}{Sen(\%)} & \multirow{2}{*}{Spe(\%)} \\  
    conditioning & conditioning   & graph-attention    & processing   
    &      &    & \\
    \cmidrule(lr){1-4}\cmidrule(lr){5-7}
    \checkmark& $\times$& $\times$& $\times$
    &$70.00\pm9.20$ &$56.27\pm10.81$ &$\textbf{99.97}\pm0.03$ \\
    \checkmark& \checkmark& $\times$& $\times$
    &$73.39\pm7.55$ &$60.96\pm9.51$ &$99.96\pm0.03$ \\
    \checkmark& \checkmark& \checkmark& $\times$
    &$\textbf{81.41}\pm6.64$ &$\textbf{81.35}\pm6.93$ &$99.84\pm0.12$ \\
    \checkmark& \checkmark& \checkmark& \checkmark
    &$81.04\pm6.50$ &$78.02\pm7.19$ &$99.88\pm0.11$ \\
    \bottomrule
\end{tabular}}
\label{tab:abla_LiVS}
\end{table*}

\begin{figure*}[t]
  \centering
  \includegraphics[width=1\textwidth]{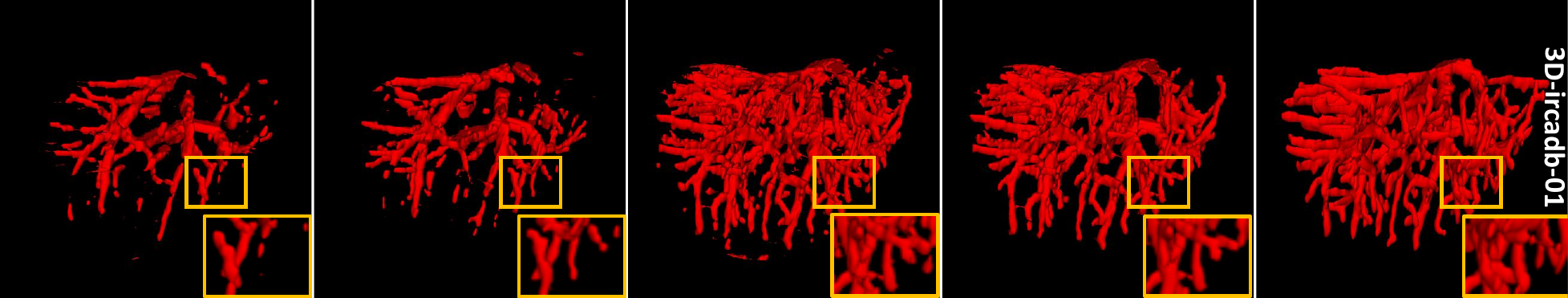} \\
  \includegraphics[width=1\textwidth]{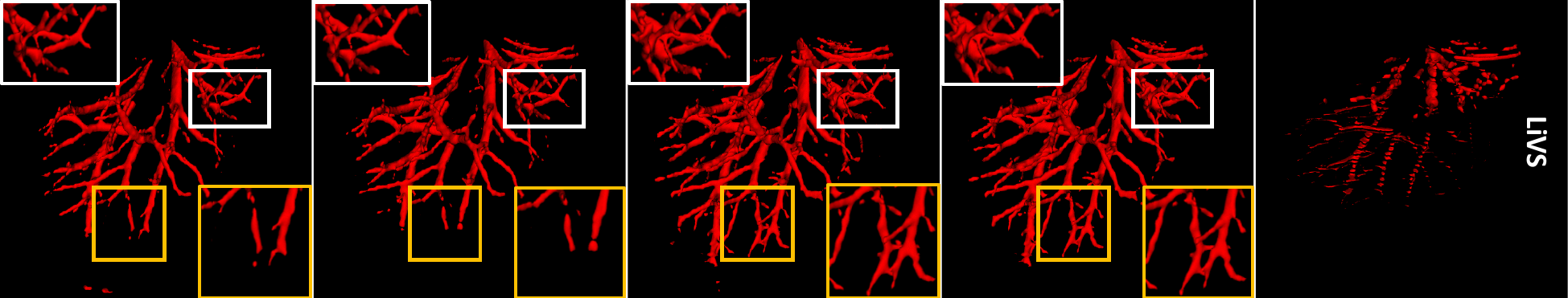} \\
  \makebox[0.2\textwidth]{\scriptsize (A) Vanilla}%
  \makebox[0.2\textwidth]{\scriptsize (B) Dynamic}%
  \makebox[0.2\textwidth]{\scriptsize (C) Dynamic multiscale }%
  \makebox[0.2\textwidth]{\scriptsize (D) Post-processing}%
  \makebox[0.2\textwidth]{\scriptsize (E) Ground truth}\\[-5px]
  \makebox[0.2\textwidth]{\scriptsize conditioning}%
  \makebox[0.2\textwidth]{\scriptsize conditioning}%
  \makebox[0.2\textwidth]{\scriptsize graph-attention}%
  \makebox[0.2\textwidth]{\scriptsize }%
  \makebox[0.2\textwidth]{\scriptsize }
  \caption{
    \update{\textbf{Visualization of the per-component ablation study for the \textsl{3D-IRCADb-01} dataset in \tab{4} and the \textsl{LiVS} dataset in \tab{abla_LiVS}.}
    Each column shows the cumulative effect of adding components (A) to (D), compared against the ground truth (E).
    (A) Vanilla conditioning;
    (B) Dynamic conditioning;
    (C) Dynamic multisclae graph-attention;
    (D) Post-processing;
    (E) Ground truth.}
  }
  \label{fig:viz_abla_cond}
\end{figure*}

\noindent\textbf{Effect of different model components and post-processing.}
For completeness, we perform the ablation studies on both the \textsl{3D-ircadb-01} and \textsl{LiVS} datasets. 
Because the three components of our model : (i) vanilla conditioning, (ii) dynamic conditioning and (iii) multiscale graph-attention, are highly interconnected (see \fig{detailed}), we cannot evaluate them independently.
To explore the influence of different conditioning levels on the performance of liver vessel segmentation, we perform an additive ablation study in \tab{4} and \tab{abla_LiVS}, \update{and visualize the ablation in \fig{viz_abla_cond}}.
We start from the \textsl{(A) vanilla conditioning model} and subsequently add new conditioning components to it. 
Noteworthy, the \textsl{(A) vanilla model} (in the first row) and the combination of the vanilla model with the \textsl{(B) dynamic conditioning model} (in the second row) achieve the highest \textsl{Spe} scores.
This is due to these models predicting fewer structures in the segmentation masks (as shown by the low \textsl{Sen} score \update{and in \fig{viz_abla_cond}}), and therefore having fewer false positives.
The final model combining all three components: \textsl{(A) vanilla conditioning}, \textsl{(B) dynamic conditioning} and \textsl{(C) multiscale graph-attention} has the highest \textsl{DSC}, and \textsl{Sen} scores.
The graph-attention conditioning contributes considerably to the performance of our model.
Post-processing is an effective way to remove noisy predictions, and to improve the \textsl{DSC} score as shown in \tab{4}, but it can negatively affect the \textsl{Sen}.
However, for datasets with discontinuous annotations (such as \textsl{LiVS}), post-processing is not mandatory because inference ensembles can also remove noisy predictions.
Overall, we conclude that all components have a beneficial effect on the vessel segmentation scores.

\begin{table*}[t]
    \centering
    \caption{
    \update{
    \textbf{Ablation study testing the effect of the number of graph nodes, on the \textsl{3D-IRCADb-01} and \textsl{LiVS} datasets.}
    No post-processing or inference ensembling was used in this ablation study on node number.
    As the number of graph nodes increase, the computations (GFLOPs) increase.
    Using $32^2\times 3$ provides a good tradeoff betwen accuracy and computations, while balancing \textsl{Sen} and \textsl{Spe}.
    }
    }
    \resizebox{1\linewidth}{!}{%
    \begin{tabular}{l ccc ccc c}
    \toprule
            \#nodes& \multicolumn{3}{c}{\textsl{3D-ircadb-01}} & \multicolumn{3}{c}{\textsl{LiVS}}& GFLOPs\\
    \cmidrule(lr){2-4}  
    \cmidrule(lr){5-7} 
            per batch& DSC(\%) & Sen(\%) & Spe(\%) &
            DSC(\%) & Sen(\%) & Spe(\%) & per batch \\
    \midrule
    $16^2\times3$ &$61.69\pm4.03$ &$76.99\pm7.95$ &$99.69\pm0.22$ &$66.13\pm10.46$ &$88.02\pm4.32$ &$99.29\pm0.29$ &$533.95$ \\ 
    $32^2\times3$ &$62.69\pm3.93$ &$79.80\pm6.33$ &$99.70\pm0.17$ &$71.90\pm8.85$ &$87.07\pm4.71$ &$99.51\pm0.20$ &$565.20$ \\ 
    $64^2\times3$ &$62.71\pm4.88$ &$54.80\pm9.00$ &$99.92\pm0.04$ &$75.57\pm5.72$ &$67.87\pm8.05$ &$99.90\pm0.05$ &$690.17$ \\
    \bottomrule
    \end{tabular}}
    \label{tab:abla_nodes}
\end{table*}
\update{
\medskip\noindent\textbf{The effect of different graph node numbers.}
We perform ablation studies testing the effect of the number of graph nodes, on \textsl{3D-ircad-01} and \textsl{LiVS} datasets.
To isolate the effect of the number of graph nodes, we do not apply post-processing or inference ensembling, in this ablation study.
In addition to the node numbers of $32{\times}32{\times}3$ per batch used in our implementation, we also report results for configurations of $16 {\times} 16 {\times} 3$ and $64 {\times} 64 {\times} 3$, as shown in \tab{abla_nodes}.
Although our model with a configuration of $64 {\times} 64 {\times} 3$ achieves the highest \textsl{DSC} and \textsl{Spe} scores on both datasets, it yields the lowest \textsl{Sen} score and incurs higher computational complexity in terms of GFLOPs.
The lower \textsl{Sen} score of our model with the $64{\times} 64{\times} 3$ configuration indicates relatively incomplete vessel segmentation, making it insufficient to meet the requirement for segmentation completeness.
The sparser node configuration of $16{\times}16 {\times} 3$ results in a lower \textsl{Spe} score, indicating a higher number of false positives in the segmentation.
To balance \textsl{Sen} and \textsl{Spe} while considering computational complexity, the $32{\times}32 {\times}3$ configuration used in our implementation represents a good tradeoff.
}

\begin{figure}[t!]
  \centering
  \includegraphics[width=.7\linewidth]{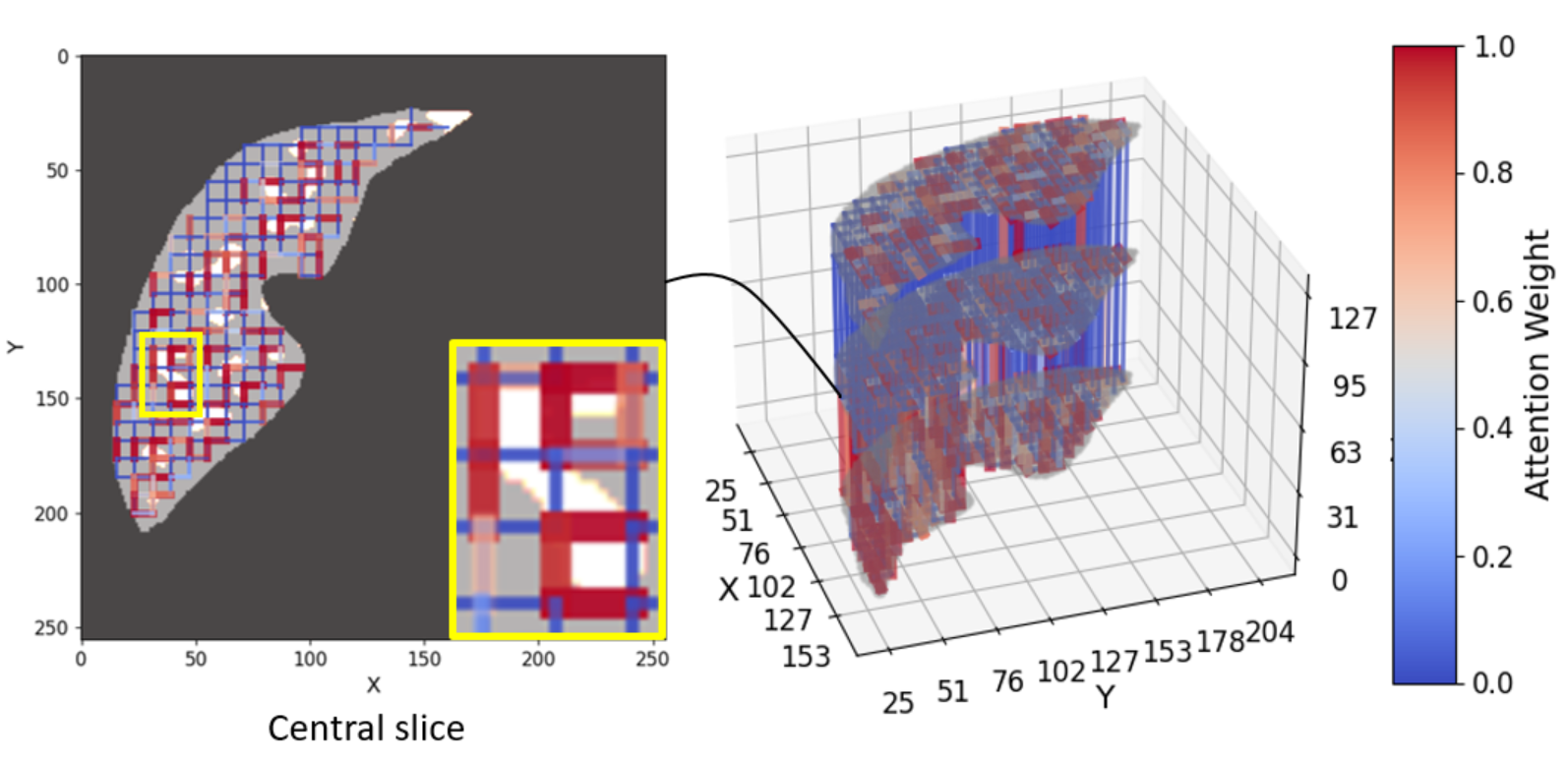}
  \caption{\textbf{Edge attention weights of a fully-connected graph in inference.}
      Blue\slash red represent low\slash high attention weights, respectively.
      We also show the box enlarged in the bottom right corner.
      The vessel area and its neighborhood attract more edge attention, thus demonstrating the utility of inputting a fully-connected graph in inference.
    }
  \label{fig:edge_atten_weights}
\end{figure}
\begin{figure}[tp!]
  \centering
  \includegraphics[width=.7\linewidth]{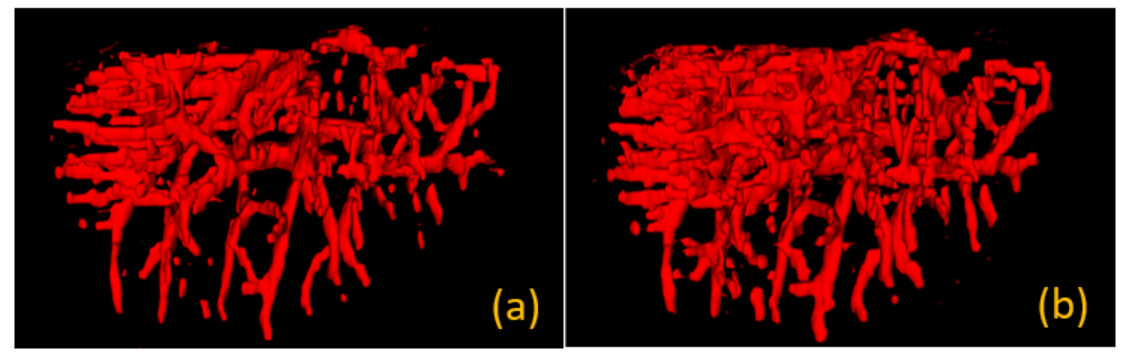}
  \caption{\textbf{Inference difference between using an empty graph, without edges (a) and a fully-connected graph (b) as the input.}
   A fully-connected graph leads to more dense and continuous predictions.
   }
  \label{fig:inference_difference_w_wo_edges}
\end{figure}

\medskip\noindent\textbf{Graph-attention in inference.}
In inference, we input a fully-connected graph in the component (C), as shown in \fig{overview}. 
While the node coordinates are uniformly distributed in the graph, the edge weights between nodes are adapted by the trained graph-attention layer. 
\fig{edge_atten_weights} visualizes the learned edge weights for an input fully-connected graph. 
In \fig{edge_atten_weights}, the cross-sectional view shows that the vessel area and its neighborhood have higher-magnitude edge-attention.
Additionally, in \fig{inference_difference_w_wo_edges} we compare the difference in predictions when inputting an empty graph, without edges (a) and a fully-connected graph (b).
Using an empty graph leads to sparser predictions.
This analysis demonstrates that graph attention layers are still effective, even with a uniform graph as input.

\begin{figure}[t!]
    \centering
      \begin{tabular}{c@{\hskip 0.01in}c@{\hskip 0.01in}c}
    \includegraphics[width=0.25\linewidth,height=0.25\linewidth]{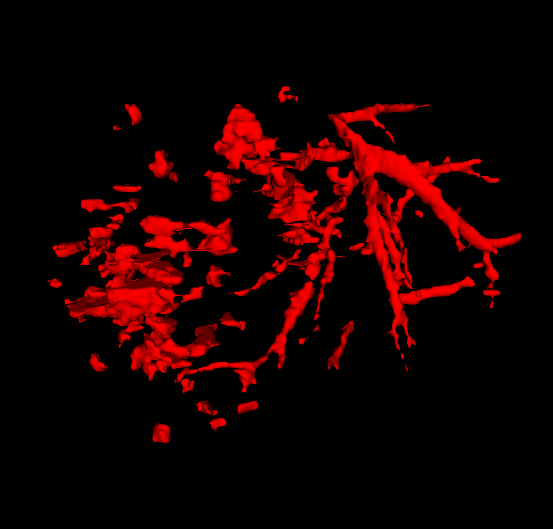} &
    \includegraphics[width=0.25\linewidth,height=0.25\linewidth]{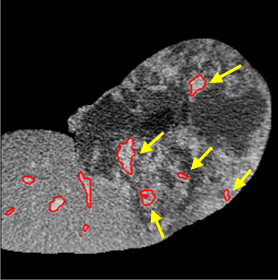} & 
    \includegraphics[width=0.25\linewidth,height=0.25\linewidth]{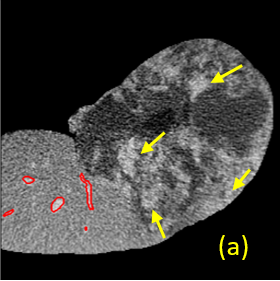} \\
    \includegraphics[width=0.25\linewidth,height=0.25\linewidth]{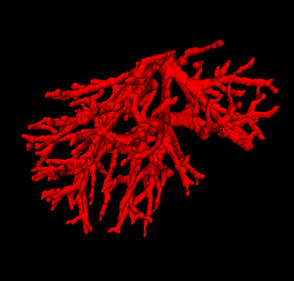} &
    \includegraphics[width=0.25\linewidth,height=0.25\linewidth]{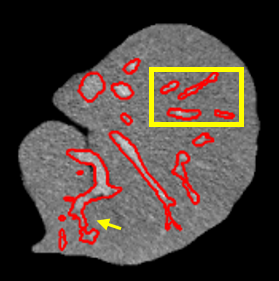} &
    \includegraphics[width=0.25\linewidth,height=0.25\linewidth]{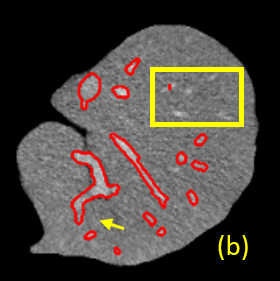} \\ 
    \scriptsize{Segmentation} &
    \scriptsize{Segmentation contours} &
    \scriptsize{Ground truth contours} \\
    \end{tabular}    
    \caption{
    \textbf{Analysis of model limitations.}
    (a) First row: Visualization of a failure case on the \textsl{LiVS} dataset.
    (b) Second row: Visualization of a failure case on the \textsl{3D-ircadb-01} dataset.
     In case (a) the failure is due to the contrast marked by yellow arrows around the tumors being misclassified as vessel structures. 
     In case (b) the failure is due to the unlabeled vessels \cite{b13,b15} (highlighted in the yellow box), causing low \textsl{DSC} scores.
     The model also makes a mistake by predicting segmentation masks where there should not be (over-segmentation), as shown by the yellow arrow. 
     This is caused by the inconsistency in annotations in the \textsl{3D-ircadb-01} dataset \cite{b40}.  
    }
  \label{fig:8}
\end{figure}

\section{Discussion}
\fig{8} shows the limitation of our model and contains examples with the worst \textsl{DSC} scores for the \textsl{LiVS} and \textsl{3D-ircadb-01} datasets.
On these worst cases, our proposed model inaccurately predicts vessel segmentations when there is a large tumor surrounded by contrast-rich regions, as seen in \fig{8}(a). 
This failure is reflected by the outliers of the \textsl{DSC} scores of the \textsl{LiVS} dataset in \fig{num_LiVS}. 
All tumors in \textsl{3D-ircadb-01} are characterized as low-intensity regions, and are not surrounded by contrast, so this failure appears only on the \textsl{LiVS} dataset. 
\fig{8}(b) shows a worst performing example on the \textsl{3D-ircadb-01} dataset. 
The low \textsl{DSC} score is due to missing annotations \cite{b13,b15}.
The yellow box highlights a case where vessels were not annotated. Our model correctly predicts this, while still being penalized in the \textsl{DSC} scores.
Using yellow arrows we indicate regions where our model over-predicts structure that is not truly present.
The inconsistency in the annotation quality in the \textsl{3D-ircadb-01} dataset\cite{b40} is an additional challenge in the  training of the model. 
This may lead to that the model learns to focus on the wrong information when predicting segmentations.

\section{Conclusions}
In this study, we focus on liver vessel segmentation from CT volumes.
To this end, we propose to augment conditional diffusion models with geometric graph-structure computed at multiple resolutions.
The role of the graph structure is to ensure connectivity in the segmentation, across neighboring slices in the CT volume. 
Moreover, we use multi-scale features in the graph, thus allowing the model to focus on small vessels, that otherwise would be missed. 
Our proposed model achieves state-of-the-art results, in terms of \textsl{DSC}, \textsl{Sen} and vessel connectivity on two standard benchmarks: 
\textsl{3D-ircadb-01} \cite{b40} and \textsl{LiVS} \cite{b41}, when compared to baselines such as \update{\textsl{HiDiff} \cite{b63}}, \update{\textsl{MERIT}\cite{b62}}, \updateSecond{\textsl{TransUNet}} \cite{b68}, \textsl{MedSegDiff} \cite{b20}, \textsl{EnsemDiff} \cite{b19}, \textsl{Swin UNETR} \cite{b28}, \textsl{nnUNet} \cite{b27} and Gao et al. \cite{b41}.

{\small
\bibliographystyle{elsarticle-num}
\bibliography{references}
}

\end{document}